\documentclass[traditabstract]{aa} 
    
\usepackage{graphicx}
\usepackage{lscape}
\usepackage{txfonts}
\usepackage{natbib}
\bibpunct{(}{)}{;}{a}{}{,}
\usepackage{url}
\usepackage[T1]{fontenc}

\newcommand{\micron}{\mbox{$\mu$m}}

\DeclareRobustCommand{\ion}[2]{%
\relax\ifmmode
\ifx\testbx\f@series
{\mathbf{#1\,\mathsc{#2}}}\else
{\mathrm{#1\,\mathsc{#2}}}\fi
\else\textup{#1\,{\mdseries\textsc{#2}}}%
\fi}

\begin{document}

   \title{CAFE: Calar Alto Fiber-fed Echelle spectrograph}

   \author{
        J.~Aceituno\inst{1}
        \and
        S.F~S\'anchez\inst{2,1}
        \and
        F. Grupp\inst{3}
        \and
        J. Lillo\inst{4}
        \and
        M. Hern\'an-Obispo\inst{5}
        \and
        D. Benitez\inst{1}
        \and
        L.M. Montoya\inst{1}
        \and
        U. Thiele\inst{1}
        \and
        S. Pedraz\inst{1}
        \and
        D. Barrado\inst{1,4}
        \and
        S. Dreizler\inst{6}
        \and
        J. Bean\inst{6}
        }

\institute{
        Centro Astron\'omico Hispano Alem\'an, Calar Alto, (CSIC-MPG),
        C/Jes\'{u}s Durb\'{a}n Rem\'{o}n 2-2, E-04004 Almer\'{\i}a,
        Spain. \email{aceitun@caha.es}.
        \and
        Instituto de Astrof\'{\i}sica de Andaluc\'{\i}a (CSIC), Camino Bajo
        de Hu\'etor s/n, Aptdo. 3004, E18080-Granada, Spain\\ \email{sanchez@iaa.es}.
        \and
        Institut fur Astronomie und Astrophysik der Universitat Munchen,
        Scheinerstr. 1, 81679 Munchen, Germany.
        \and
        Departamento Astrof\'isica, Centro de Astrobiolog\'ia (INTA-CSIC), ESAC campus, P.O. Box
        78, E-28691, Villanueva de la Cañada, Spain.
        \and
        Universidad Complutense de Madrid, Av. Complutense s/n, 28040 Madrid,
        Spain.
        \and
        University of Goettingen, Wilhelmsplatz 1  37073
        Goettingen, Germany.
        }

   \date{Received ----- ; accepted ---- }


  \abstract{
We present here CAFE, the Calar Alto Fiber-fed Echelle
spectrograph, a new instrument built at the Centro Astronomico
Hispano Alem\'an (CAHA). CAFE is a single fiber, high-resolution
($R\sim$70000) spectrograph, covering the wavelength range between
3650-9800\AA. It was built on the basis of the common  design
for Echelle spectrographs. Its main aim is to measure radial
velocities of stellar objects up to $V\sim$13-14 mag with a
precision as good as a few tens of $m~s^{-1}$. To achieve this
goal the design was simplified at maximum, removing all possible
movable components, the central wavelength is fixed, so the wavelentgth  coverage; no filter wheel, one slit and so on, with a particular care taken
in the thermal and mechanical stability. The instrument is fully operational and
publically accessible at the 2.2m telescope of the Calar Alto
Observatory.

In this article we describe (i) the design, summarizing its
manufacturing phase; (ii) characterize the main properties of the
instrument; (iii) describe the reduction pipeline; and (iv) show
the results from the first light and commissioning runs.  The
preliminar results indicate that the instrument fulfill the
specifications and it can achieve the foreseen goals. In
particular, they show that the instrument is more efficient than
anticipated, reaching a $S/N\sim$20 for a stellar object as faint
as $V\sim$14.5 mag in $\sim$2700s integration time. The instrument is a wonderful machine for exoplanetary research (by studying large samples of possible systems cotaining massive planets), galactic dynamics (high precise radial velocities in moving groups or stellar associations) or astrochemistry.

}

   \maketitle


\section{Introduction}

The Calar Alto Fiber-fed Echelle spectrograph (CAFE) is an
instrument manufactured at the Centro Astron\'omico Hispano
Alem\'an (CAHA) to replace FOCES (Pfeiffer 1992, Pfeiffer et al.
1998), the high-resolution echelle spectrograph at the 2.2m
telescope of the observatory that was being operated during the
period 1997-2010. CAFE was designed following the common 
concept of this kind of instrument e.g. Kaufer and Pasquini 1998,
Stahl et al. 1999, Raskin et al. 2010, and therefore its design is
very similar to that of FOCES e.g. Pfeiffer et al. 1992, Pfeiffer
et al. 1994, Pfeiffer et al. 1998.  The instrument was designed to
achieve an spectral resolution of $R\sim 70000$, covering the
wavelength range between 3850-9800\AA. The main improvements of
our design were focused on the increase of the stability and the
sensitivity as much as possible. In particular, (i) we design a
new camera, improving the one from FOCES; (ii) we use new branch,
highly efficient and long-term stable fibers; (iii) the 
instrument has been located in an isolated room, thermalized and
stabilized against vibrations (as we illustrate latter);(iv) most
of the possible mobile parts in this kind of instruments have been
substituted by fixed elements, to increase the stability of the
system; and finally (v) a new more efficient CCD, with a smaller
pixel has been acquired.

We expected that these improvements increase the efficiency and
quality of the data with respect to its predecessor. The ultimate
goal is that CAFE would achieve precisions of $\sim$10-20
$m~s^{-1}$ in the measurement of radial velocity of stellar
objects down to $V\sim$14 magnitudes. We are searching for the maximun radial velocity accuracy for a simple
  and unexpensive, fast track instrument. The goal was to produce a
  competitive instrument for several key areas such exoplanetary searches,
  including the Kepler candidates, and the exploitation of GAIA \cite{per2003}. Thus,the $\sim$10-20
$m~s^{-1}$ is a compromise, and allows detailed studies of large samples of
  moderately faint stars. However, we would like to note that this instrument
  is built foreseen a single science case, but not to provide the community of
users of the observatory with a facility to cover the same studied that were
performed during the last 10 years using FOCES, improving the performance when
feasible. 

 The intrument operates in a single mode, (as
FEROS at ESO 1.52m \cite{kau1997}, HARPS at ESO 3.6m
\cite{pep2000}) without any possible adjustment to increase either
the resolution or the efficiency (to the contrary of other similar
instruments with more movable elements, eg., FOCES, SOFIN at NOT
2.6m telescope \cite{pro1995}). 

The distribution of this article is the following one: In Section
\ref{sec} we describe the main properties of the instrument. The
details of the design and manufacturing are given in Section
\ref{con}.  The results from the tests performed during the First
Light are shown in Section \ref{first}.  Finally, the tests
performed during the Commissioning and their results are shown in
Section \ref{comm}.  A summary of the main characteristics of the
instrument are included in Section \ref{sum}.

\section{Main properties of the instrument}
\label{sec}

CAFE is a stationary echelle spectrograph located at a remote
laboratory in the dome building of the 2.2m telescope. This room
is located below the main telescope structure, separated from the
rest of the building, in order to reduce the effects of any
mechanical vibration. The mechanical stability is increased by
sophisticated pneumatic stabilization system installed in the
optical bench manufactured by NewPort Corporation (model
I-2000-428). This guarantees a much better stability of the
instrument, increasing its performance and accuracy for radial
velocity measurements of fainter objects. The room has thick
reinforced concrete walls, as been part of the main support of the
telescope itself. This also provides a good thermal-isolation to
the room. The instrument itself is separated $\sim$18m from the
Cassegrain focus of the telescope. The light is conducted by a
single fiber and coupled to the focal plane with an improved
version of the FOCES telescope module. It was manufactured by the
company Ceramoptec and it has an stainless steel tube as outer
protection and ETFE (Ethylene tetrafluoroethylene) with Kevlar for
strain relief as inner protection tube.

The optical design of the instrument camera has been optimized,
based on the knowledge acquired with FOCES. Based on modern
optical software a better PSF could be achieved over the field of
view. The system has been optimized for the size of the CCD in
use. Finally, we equipped the instrument with a new CCD camera, an
iKon-L of Andor Technologies company, with 2048x2048 pixels of
13.5 mu. This CCD has a better quantum efficiency, lower
readout-noise and higher read-out speed than the currently used by
FOCES. We expect to increase the efficiency by at least a
$\sim$10\% using this detector.

As indicated before, the echelle image covers the visible spectral
region from 3960 to 9500 nm, distributed in 84 orders. They are
displayed in 84 spectral orders with full spectral coverage.
Spectral orders are separated by 20 pixels in the blue and 10
pixels in the red. The maximum expected spectral resolution is R =
$\lambda/\Delta\lambda$ = 67000 with a 2 pixel resolution element.
Table \ref{tab:main} summarizes the main properties of the
instrument and Table \ref{tab:elements} shows their references and
manufacturers.

\begin{table}
\caption{Basic features of CAFE.}
\label{tab:main}      
\begin{center}
\begin{tabular}{ll}\hline\hline
Design & Echelle spectrograph\\
Telescope & Calar Alto 2.2m\\
Resolution & 62000$\pm$5000 A\\
Wavelength & 3960 - 9500 A\\
Sensitivity & SNR $\sim$30 mag 14.5 and 2700sec\\ \\
TELESCOPE MODULE\\ \hline
Calibration lamps & Hal and ThAr\\
Entrance diaphragm & 2.4 arcseconds (200$\mu$m)\\  \\ OPTICAL FIBER\\
\hline
Type & Polymicro FBP100140170\\
Length & 17.5m\\
Inner protection tube & ETFE with Kevlar for strain relief\\
Outer protection tube & stainless steel tube\\
Micro-lenses & N-F2, Both ends\\ \\SPECTROGRAPH\\  \hline Optical
bench & 2400 x 1200 x 203mm\\
Entrance slit width & 100$\mu$m\\
Grating & 31.6 g/mm Blaze angle=63.9 degrees\\
Collimators & OAP1 $\lambda$/20 FL=60.0 D=10.0 OAD=7.0\\
   & OAP2 $\lambda$/20 FL=60.0 D=10.0 OAD=9.0\\

Prisms & LF5, Deviation angle 33 degrees\\
Camera & f/3\\
CCD back illuminated& 2048x2048 pixels, 13.5$\mu$m\\

\hline 
\end{tabular}
(*) OAP: Off axis parabola, FL: focal length, OAD: off axis distance (respect
to the center of the mirror.These magnitudes are in inches.
\end{center}
\end{table}

\begin{table}
\caption{Summary of the main elements of CAFE}
\label{tab:elements}      
\begin{center}
\begin{tabular}{lll}        
\hline\hline                 
Item & Manufacturer & Reference  \\
\hline
CCD & Andor Technologies  & iKON-L DZ936NBV \\
CCD cooling system & Solid State & XW-CHIL-150 \\
Pneumatic isolators & NewPort Corp &  I-2000-428 \\
Collimators & NewPort Corp & OAP60-02-10Q\\
  &  & OAP60-04-10SQ \\
Flat mirror & Bernhard Halle Nachff & -- \\
Grating & NewPort Corp & 53044ZD06-411E \\
Dispersion prisms & Prazisionsoptics Gera & 10586.09.001\\
Custom mounts & Indicam technologies & -- \\
Fiber optics & Ceramoptec & -- \\
Camera optics & Prazisionsoptics Gera & -- \\

\hline
\end{tabular}
\end{center}
\end{table}


\section{Detailed description of the instrument}
\label{con}

\subsection{Mechanical and thermal stability}

As indicated before, the spectrograph is located inside a
controlled thermal environment in order to minimize any possible
thermal drifts during an observing night.

\begin{figure}
\resizebox{\hsize}{!}
{\includegraphics[width=\hsize]{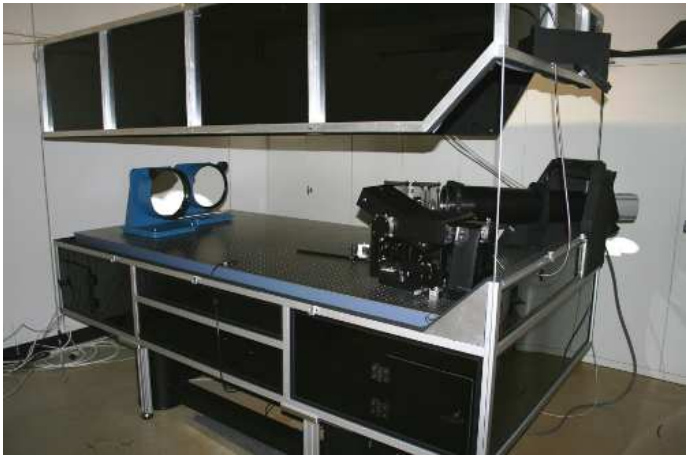}}
  \caption{\label{CAFE2}
 The figure shows a view of an CAFE spectrograph with the cabinet opened.
 The internal optical elements are seen on the optical
 bench. The upper side is as straight as a die to avoid any small
 apertures that can break the thermal isolation. Due to that,
 the upper part has to be lift up with the help of a tackle.
 }
  \end{figure}

The upper part of the cabinet can be lifted up with a tackle
anchored on the roof of the room, as shown if Fig.\ref{CAFE2},
getting access to the optical bench. The enclosure is kept closed
during the observations.

No electronic/mechanical control system is needed once the
instrument is integrated, due to its design. The only active
elements inside the cabinet are the focus stage of the camera and
the shutter, which are kept disconnected during a normal
operation. These issues add a better stability of the mechanical
mounts installed on the optical bench.

Following the same philosophy, the entrance slit width  of the
spectrograph is fixed to 100 $\mu$m. The maximum expected
spectrograph resolution for a 2.2m telescope is R$\phi$ = 60300
with the slit width $\phi$ entered in arcseconds; at the 2.2 m
telescope the standard slit width corresponds to 100 $\mu$m which
subtends a 1.2 arcseconds angle on the sky. Thus a CCD chip with
2x2k pixels of 13.5 $\mu$m distance would yield a 2 pixel
resolution of R = 72000 with increased spectral coverage. The
selection of the slit's width matches with the average values of
seeing at the Calar Alto observatory which corresponds to
0.89arcseconds \cite{san2007}.

\subsection{Optical design}

\begin{figure}
\resizebox{\hsize}{!}
{\includegraphics[width=\hsize]{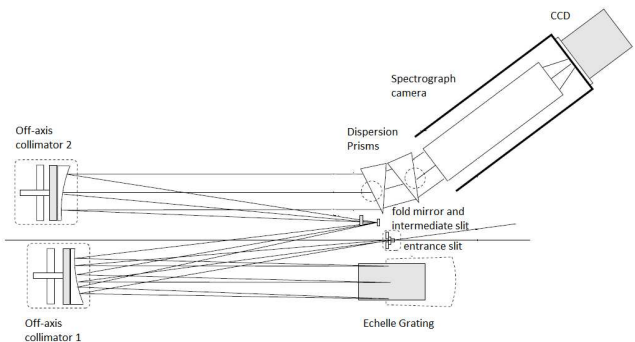}}
 \caption{\label{fig:Opt}
 Optical layout of the CAFE spectrograph with entrance slit, collimators, Echelle grating, folding mirror, prism/grism
cross-disperser, and camera.
}
 \end{figure}

The optical design of CAFE follows a white pupil concept, similar
in many aspects to the one its precedent, FOCES (Baranne 1998).
The optical layout is shown in Fig.\ref{fig:Opt}.

\begin{figure}
\resizebox{\hsize}{!}
{\includegraphics[width=\hsize]{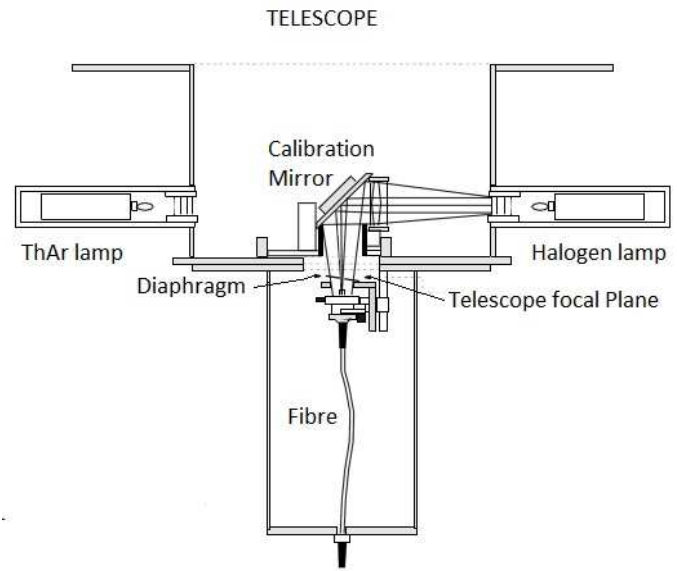}}
  \caption{\label{Telmod}
 The figure shows the telescope module. It is attached to the
 Cassegrain focus of the telescope and contains the calibration
 lamps, calibration mirror, a diaphragm and the fiber that brings
 the light to the spectrograph.
 }
  \end{figure}

The light is collected by the telescope module (see Fig.\ref{Telmod}), which is attached to the Cassegrain focus of the
telescope . It comprises a Halogen and Thorium-Argon lamps that
are used as a continuum and wavelength calibration respectively,
and an optical fiber that feeds the spectrograph. A motorized
device allows to place any of the calibrations lamps in the
optical path when they are needed. Their beams are designed to
have roughly the same $f$-ratio ($f/8$) as the telescope beam to
illuminate the entrance aperture with the same light cone. A
45$\,^{\circ}{\rm }$ tilted mirror reflects the light into the
fiber head and it might be rotated to accept light coming from any
of the comparison lamps sources. Currently halogen and Thorium
Argon lamps are available for flat field and wavelength
calibration respectively.

Light entering into the fiber passes through a circular diaphragm
which is located in a small tilted mirror just atop the fiber
head. The diameter of such diaphragm is fixed to 200 $\micron$
that determines the angular field accepted by the fiber and
corresponds to an sky area of 2.4arcseconds. This small tilted
mirror also allows guiding capabilities on the entrance aperture
using the telescope guiding facilities.

\begin{figure}
\resizebox{\hsize}{!}
{\includegraphics[width=\hsize]{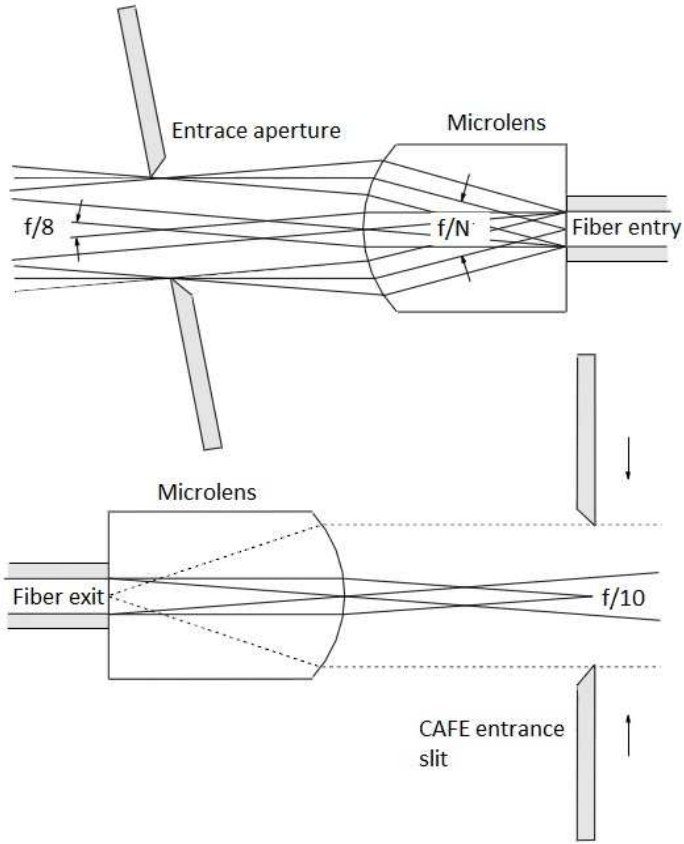}}
  \caption{\label{Microlenses}
Principle of working of the microlenses glued to the entrance and
exit of the fiber. Upper panel: Fiber feed at the telescope
module. Lower panel: Entrance at the spectrograph (\cite{pfe1998}).
 }
  \end{figure}

Both ends of the fiber have a microlens glued to each surface and
the corresponding principle is shown is Fig \ref{Microlenses}. As
in the case of FOCES, a polymicro fiber was selected to be used
with CAFE because it has been reported to have the least
degradation among different available ones, e.g. \cite{cra2008,avi2006,avi1988}.

A Fourier coupling is used. That means that the radial
distribution of light in the telescope focal plane is transferred to an angular
distribution to the fiber. As angular distribution of light couple to the
fiber is presevered by the fiber (modulo FDR), the Fourier lens at the end
recovers the radial information from the angular one.

The first element in the optical bench is the entrance slit used
to recover the original resolution when the entrance diaphragm at
the Cassegrain focus is widely open to let pass starlight in case
of bad seeing.  It is immediately aside of the folding mirror so
the entrance slit and its spectral image are therefore very nearby
as close as 0.9mm. 

To follow a white pupil design, the spectrograph itself is
collimated with two large off-axis parabolic mirrors. The 15 cm
beam leaving the 31.6 lines/mm R2 echelle is refocussed in the
vicinity of a small folding mirror, used to reflect the converging
beam in the intermediate slit image which passes a very efficient
straylight baffle.

The cross-dispersion is achieved with two LF5 prisms installed on
a symmetric tandem mounting which is under manual control. Instead
of a low-order grating, a double prism for cross-dispersion is
used, which accounts for a less strongly changing inter order
distance, and it significantly reduces local straylight in the
spectrum.

Finally, the beam is imaged with an f\/3 transmission camera onto
a field centered on a back-illuminated CCD with 13.5 $\mu$m pixel
size. The optical design of the instrument camera has been
optimized, based on the knowledgement acquired with FOCES. Based
on modern optical software a better PSF could be achieved over the
field of view. The system has been optimized for the size of the
CCD in use.

\begin{figure}
\begin{center}
\includegraphics[width=\hsize, clip=true,trim=1 200 1 1]{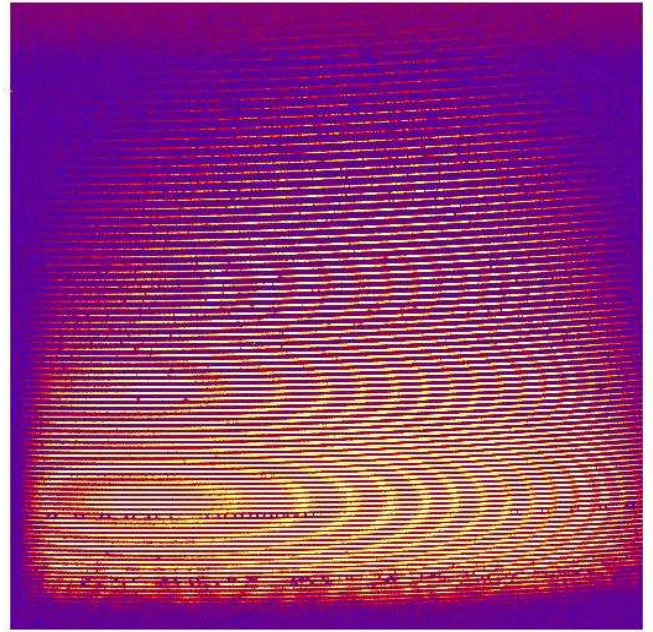}
\caption{Raw data of the first science frame taken with CAFE, on
the star HR4728. The orders are displayed from red, at the bottom,
to blue, a the top. For each order, the wavelength range runs from
blue, at the left, to red, at the right. \label{fig:HR4728raw}}
\end{center}
\end{figure}

\subsection{Pipeline}
\label{sec:pipe}

CAFE has been delivered with an automatic reduction pipeline. The
main goal of this pipeline is to provide the user with a simple
reduction of the data that can be used to test their quality
during the observing run. Depending of the particularities of the
science case, the data reduced using the pipeline could be used
for science purpose or not, although this was not the main purpose
of this package. The software is available for the community on
request.

The reduction pipeline is based on R3D \cite{san2006}. Although
the package was not originally designed to reduce Echelle data,
these data share some of the basic properties of any fiber-fed
spectrograph. The pipeline requires four basic entries to perform
the reduction: (1) the science raw-frame, (2) a bias frame
acquired during the night, (3) a continuum lamp frame, and (4) an
arc-lamp frame (ThAr, in this case). The main reduction steps
performed are the following: (i) the raw frames are corrected by
the corresponding bias; (ii) the location of the orders across the
cross-dispersion axis is determined based on the intensity peak of
each order at the central pixel of the CCD. Then, each order is
traced along the dispersion axis, using a similar algorithm; (iii)
the science spectrum corresponding to each order is then extracted
using a Gaussian extraction algorithm \cite{san2006,san2012}. A
similar procedure is used to extract the continuum and arc lamp
spectra: (iv) The continuum extracted spectra are then used to
create a master flat. For doing so, a low-order polynomial
function is fitted to each continuum extracted spectrum, which is
then divided by this smooth curve. The science frames are divided
by this latter flat frame, correcting both the effects of the
fringing and normalizing the transmission order-by-order. (v)
Then, the wavelength calibration is derived for each order using
the extract arc-lamp spectra. The identified emission lines are
stored in an internal look-up-table (LUT)\footnote{a summary of
this LUT is shown in
http://www.caha.es/CAHA/Instruments/CAFE/cafe/orders2.pdf}. It is
used a five-order polynomial function to derive the wavelength
calibration order-by-order. (vi) The wavelength calibration is
applied to the science frames, normalizing the wavelength step to
the nominal one order-by-order as shown in Table \ref{tab:orders}.
(vii) Finally, a rough flux calibration is performed using a
master-transmission curve derived during the commissioning runs.

\begin{table*}[t]
\caption{CAFE: List of the clearly detected orders, including the
order, the wavelength range and the spectral sampling (after
re-sampling).}
\label{tab:orders}      
\begin{center}
\begin{tabular}{rlll|rlll|rlll}        
\hline\hline                 
\# & $\lambda_{\rm start}$ & $\lambda_{\rm end}$ &
$\Delta\lambda/pix$ & \# & $\lambda_{\rm start}$ & $\lambda_{\rm
end}$ & $\Delta\lambda/pix$ &
\# & $\lambda_{\rm start}$ & $\lambda_{\rm end}$ & $\Delta\lambda/pix$ \\
      & (\AA)  & (\AA)  & (\AA/pix) &
      & (\AA)  & (\AA)  & (\AA/pix) &
      & (\AA)  & (\AA)  & (\AA/pix) \\
\hline                                   
60 & 9432.964 & 9558.496 & 0.0645  & 88  & 6430.853 & 6516.885 & 0.0442  & 116 & 4877.880 & 4943.502 & 0.0337  \\
61 & 9278.291 & 9401.792 & 0.0634  & 89  & 6358.568 & 6443.648 & 0.0437  & 117 & 4836.160 & 4901.236 & 0.0334  \\
62 & 9128.607 & 9250.142 & 0.0624  & 90  & 6287.889 & 6372.039 & 0.0432  & 118 & 4795.146 & 4859.686 & 0.0331  \\
63 & 8983.674 & 9103.305 & 0.0614  & 91  & 6218.764 & 6302.003 & 0.0427  & 119 & 4754.821 & 4818.833 & 0.0328  \\
64 & 8843.271 & 8961.057 & 0.0605  & 92  & 6151.141 & 6233.490 & 0.0423  & 120 & 4715.168 & 4778.662 & 0.0326  \\
65 & 8707.188 & 8823.185 & 0.0596  & 93  & 6084.972 & 6166.449 & 0.0418  & 121 & 4676.170 & 4739.154 & 0.0323  \\
66 & 8575.229 & 8689.491 & 0.0587  & 94  & 6020.210 & 6100.835 & 0.0414  & 122 & 4637.811 & 4700.294 & 0.0321  \\
67 & 8447.208 & 8559.787 & 0.0578  & 95  & 5956.812 & 6036.603 & 0.0410  & 123 & 4600.075 & 4662.066 & 0.0318  \\
68 & 8322.953 & 8433.897 & 0.0570  & 96  & 5894.734 & 5973.708 & 0.0405  & 124 & 4562.947 & 4624.454 & 0.0316  \\
69 & 8202.299 & 8311.657 & 0.0561  & 97  & 5833.936 & 5912.109 & 0.0401  & 125 & 4526.413 & 4587.443 & 0.0313  \\
70 & 8085.092 & 8192.908 & 0.0554  & 98  & 5774.378 & 5851.768 & 0.0397  & 126 & 4490.459 & 4551.020 & 0.0311  \\
71 & 7971.187 & 8077.504 & 0.0546  & 99  & 5716.023 & 5792.646 & 0.0393  & 127 & 4455.070 & 4515.170 & 0.0308  \\
72 & 7860.446 & 7965.305 & 0.0538  & 100 & 5658.835 & 5734.706 & 0.0389  & 128 & 4420.234 & 4479.881 & 0.0306  \\
73 & 7752.738 & 7856.180 & 0.0531  & 101 & 5602.779 & 5677.913 & 0.0386  & 129 & 4385.938 & 4445.138 & 0.0304  \\
74 & 7647.942 & 7750.003 & 0.0524  & 102 & 5547.822 & 5622.233 & 0.0382  & 130 & 4352.168 & 4410.929 & 0.0301  \\
75 & 7545.939 & 7646.658 & 0.0517  & 103 & 5493.932 & 5567.634 & 0.0378  & 131 & 4318.914 & 4377.243 & 0.0299  \\
76 & 7446.621 & 7546.032 & 0.0510  & 104 & 5441.077 & 5514.086 & 0.0375  & 132 & 4286.164 & 4344.066 & 0.0297  \\
77 & 7349.883 & 7448.020 & 0.0504  & 105 & 5389.229 & 5461.557 & 0.0371  & 133 & 4253.905 & 4311.388 & 0.0295  \\
78 & 7255.624 & 7352.520 & 0.0497  & 106 & 5338.359 & 5410.019 & 0.0368  & 134 & 4222.128 & 4279.198 & 0.0293  \\
79 & 7163.752 & 7259.438 & 0.0491  & 107 & 5288.440 & 5359.444 & 0.0364  & 135 & 4190.821 & 4247.485 & 0.0291  \\
80 & 7074.176 & 7168.682 & 0.0485  & 108 & 5239.445 & 5309.805 & 0.0361  & 136 & 4159.974 & 4216.237 & 0.0289  \\
81 & 6986.812 & 7080.168 & 0.0479  & 109 & 5191.348 & 5261.078 & 0.0358  & 137 & 4129.577 & 4185.446 & 0.0287  \\
82 & 6901.579 & 6993.811 & 0.0473  & 110 & 5144.125 & 5213.236 & 0.0355  & 138 & 4099.620 & 4155.101 & 0.0285  \\
83 & 6818.399 & 6909.536 & 0.0468  & 111 & 5097.753 & 5166.255 & 0.0352  & 139 & 4070.093 & 4125.192 & 0.0283  \\
84 & 6737.199 & 6827.267 & 0.0462  & 112 & 5052.209 & 5120.114 & 0.0348  & 140 & 4040.988 & 4095.710 & 0.0281  \\
85 & 6657.910 & 6746.933 & 0.0457  & 113 & 5007.470 & 5074.789 & 0.0345  & 141 & 4012.296 & 4066.645 & 0.0279  \\
86 & 6580.464 & 6668.467 & 0.0452  & 114 & 4963.516 & 5030.259 & 0.0342  & 142 & 3984.007 & 4037.990 & 0.0277  \\
87 & 6504.799 & 6591.805 & 0.0447  & 115 & 4920.326 & 4986.504 & 0.0340  & 143 & 3956.113 & 4009.736 & 0.0275  \\
\hline
\end{tabular}
\end{center}
\end{table*}

\section{Commissioning}
\label{comm}

\subsection{First light}
\label{first}

The first light of CAFE took place the night of the 24th of May
2011, when the instrument was for the first time installed in the
telescope. The first observed science target was a bright ($V\sim$6 mag)
star, HR4728, selected by visibility and luminosity.The early tests performed along this night were focused on the
identification of the orders and wavelength range covered by each
one. The identification of the orders was a fundamental and not
trivial task prior to the proper extraction and wavelength
calibration of each frame. 

\begin{figure}
\begin{center}
\includegraphics[width=6.7cm,angle=270]{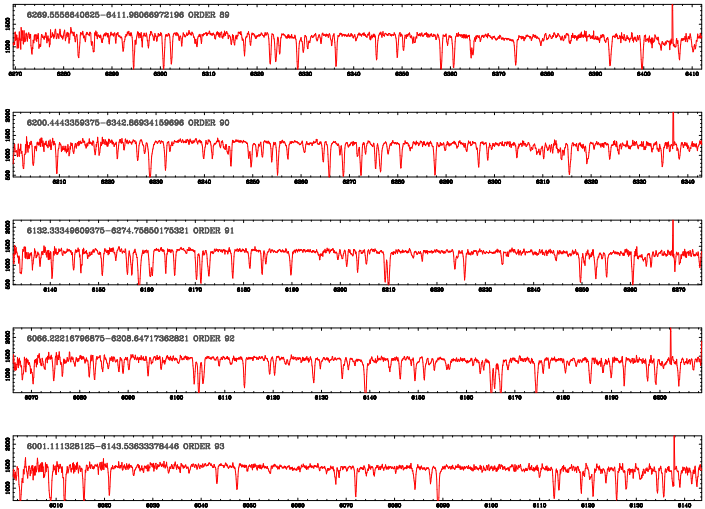}
\caption{Detail of a few orders of the extracted spectra of
HR4728, reduced using the early version of the
pipeline.\label{fig:HR4728redu}}
\end{center}
\end{figure}

\begin{figure*}
\begin{center}
\includegraphics[height=12cm,angle=270]{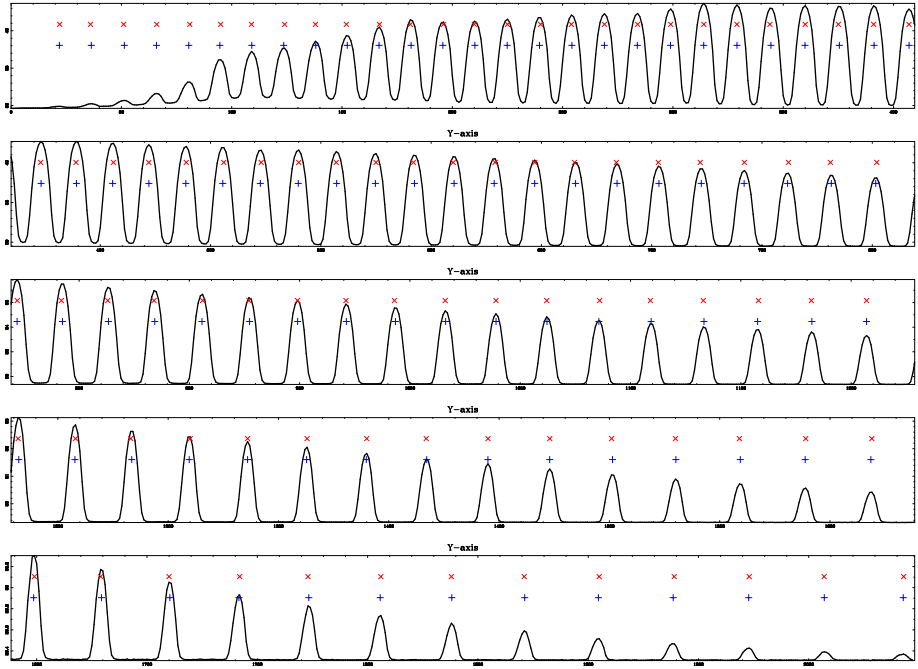}
\caption{Vertical section of a continuum exposure at the central
  X-axis pixel, showing the projection on the CCD of each order, from
red wavelengths (top-left) to blue ones (bottom-right). The
location of the peak intensity of each order is marked with a red
cross (central pixel), and with a blue cross (centroid).
\label{fig:peak_find}}
\end{center}
\end{figure*}

First, the location of the projection of each order in the CCD, as
seen in Fig.\ref{fig:HR4728raw} is determined using the
corresponding tracing routines. Fig.\ref{fig:peak_find} shows a
vertical cut of continuum lamp exposure used to trace these
locations. The distribution of flux along the cross-dispersion
axis corresponding to each of the orders is shown. For each one,
the location of the peak intensity is indicated with a red
(central pixel) and blue (hyperbolic centroid) cross.

The identification of the orders was done after tracing and
extracting a ThAr frame (following the reduction steps explained in section
\ref{sec:pipe}).

Table \ref{tab:peaks} shows the 84 detected orders, including the
Y-coordinate of the peak intensity of each order projected in the
CCD at the central pixel in the X-axis, i.e., the pixel marked
with a red cross in Fig.\ref{fig:peak_find}. Table
\ref{tab:orders} shows, for each order, the wavelength range
sampling, once normalized to a common spectral sampling per pixel.

\begin{table}[t]
\caption{CAFE: List of detected orders.}
\label{tab:peaks}      
\begin{center}
\begin{tabular}{rl|rl|rl}        
\hline\hline                 
\# & Y pixel & \# & Y pixel &
\# & Y pixel $^*$\\
\hline
60 & 152 & 88 & 593 & 116 & 1182 \\
61 & 167 & 89 & 611 & 117 & 1207 \\
62 & 181 & 90 & 629 & 118 & 1232 \\
63 & 196 & 91 & 647 & 119 & 1258 \\
64 & 210 & 92 & 666 & 120 & 1283 \\
65 & 225 & 93 & 685 & 121 & 1310 \\
66 & 240 & 94 & 703 & 122 & 1336 \\
67 & 254 & 95 & 723 & 123 & 1363 \\
68 & 269 & 96 & 742 & 124 & 1390 \\
69 & 284 & 97 & 762 & 125 & 1417 \\
70 & 299 & 98 & 782 & 126 & 1445 \\
71 & 314 & 99 & 802 & 127 & 1473 \\
72 & 329 & 100 & 822 & 128 & 1501 \\
73 & 345 & 101 & 843 & 129 & 1530 \\
74 & 360 & 102 & 864 & 130 & 1559 \\
75 & 376 & 103 & 884 & 131 & 1589 \\
76 & 391 & 104 & 906 & 132 & 1618 \\
77 & 407 & 105 & 927 & 133 & 1649 \\
78 & 423 & 106 & 949 & 134 & 1679 \\
79 & 439 & 107 & 971 & 135 & 1710 \\
80 & 456 & 108 & 994 & 136 & 1741 \\
81 & 472 & 109 & 1016 & 137 & 1773 \\
82 & 489 & 110 & 1039 & 138 & 1805 \\
83 & 506 & 111 & 1062 & 139 & 1837 \\
84 & 523 & 112 & 1086 & 140 & 1870 \\
85 & 540 & 113 & 1109 & 141 & 1905 \\
86 & 557 & 114 & 1133 & 142 & 1938 \\
87 & 575 & 115 & 1158 & 143 & 1972 \\
\hline
\end{tabular}

(*) For each order we show the y-axis pixel at the x-axis center
pixel in the CCD. The data showed here were obtained during the
commissioning. An updated table can be found at
http://w3.caha.es/CAHA/Instruments/CAFE/index.html

\end{center}
\end{table}

\subsection{Efficiency of the Instrument}

The net efficiency of the instrument was derived by comparing the
expected photons to obtain from a certain source with the real
number of photons acquired by the instrument.  For doing so, it
was used the exposures on the calibration stars during the night
of July 17th of 2011. We observed 5 different calibration stars
along that night. The grammes were reduced using the pipeline
(described latter), and finally it was extracted a
(flux)-uncalibrated spectra for each of the orders, in counts.
Then, counts were transformed into photoelectrons using the gain
of the CCD for each wavelength ($n_{e,det}$).

\begin{figure}
\begin{center}
\includegraphics[angle=270, width=\hsize]{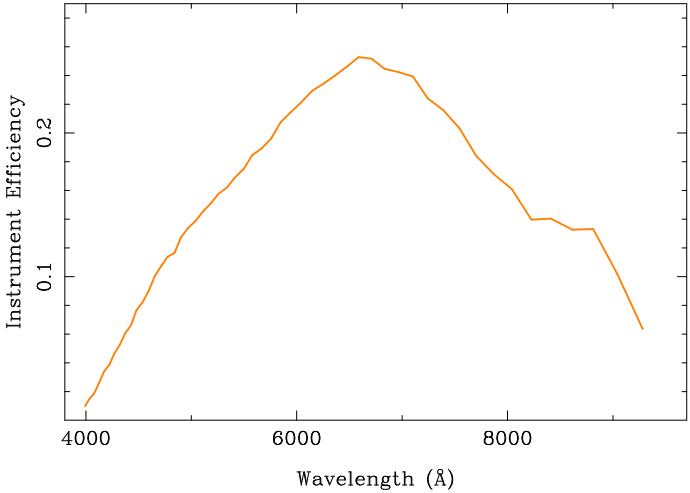}
\caption{Overall CAFE efficiency (including telescope) derived
from the analysis of the calibration star taken during the
commissioning run. \label{fig:eff}}
\end{center}
\end{figure}

To derive the number of expected photoelectrons it is required to
use the known flux-calibrated spectra of the considered standard
star. The observed spectrophotometric standard stars were all
extracted from the Oke (1990) catalogue \footnote{
  http://www.caha.es/pedraz/SSS/Oke/oke.html}, which flux density is
provided in units of 10$^{-16}$ Erg s$^{-1}$ \AA$^{-1}$ cm$^{-2}$.
The amount of flux ($F$) at a certain wavelength $\lambda$, in a
wavelength interval $\Delta \lambda$, of a star (or any other
target), with flux density $f_\lambda$, collected by a telescope
of collecting area ($\Delta S$), in a time interval ($\Delta t$)
is given by the formula:

$$  F = f_\lambda\cdot \Delta S\cdot \Delta t\cdot \Delta \lambda $$

On the other hand, the energy of a single photon is:

$$ f_{photon} = \frac{hc}{\lambda}$$

The ratio between both quantities, gives the number of {\it
expected} photoelectrons ($n_{e,exp}$). This number can be
compared directly with the number of {\it detected}
photoelectrons, obtained from the reduced data as described
before. Finally, the net efficiency of the instrument, as a
function of the wavelength, is defined as:

$$ efficiency = \frac{n_{e,det}}{n_{e,exp}} $$

Fig.\ref{fig:eff} shows the derived efficiency of the
instrument (plus telescope and detector), as a function of the
wavelength. CAFE is significantly more efficient than FOCES, at
any wavelength range \cite{pfe1998}. In comparison with other
similar Echelle spectrographs, available at telescopes of a
similar aperture, CAFE has a similar peak efficiency as FEROS
(\cite{gue1999}, \cite{kau1999}), although this instrument is more
efficient in the blue end. The main difference seems to be the
efficiency of the CCD, which for FEROS has a coating whose
efficiency changes from blue to red from one side to another
across the CCD (ie., its coating is optimized for the wavelength
range covered by each order). Therefore, a simple way to improve
even more the efficiency of the instrument would be to acquire a
similar CCD, although is not considered in the near future.

\subsection{Instrumental Focus and Wavelength Resolution}
\label{sec:focus}

\begin{figure}
\begin{center}
\includegraphics[width=\hsize]{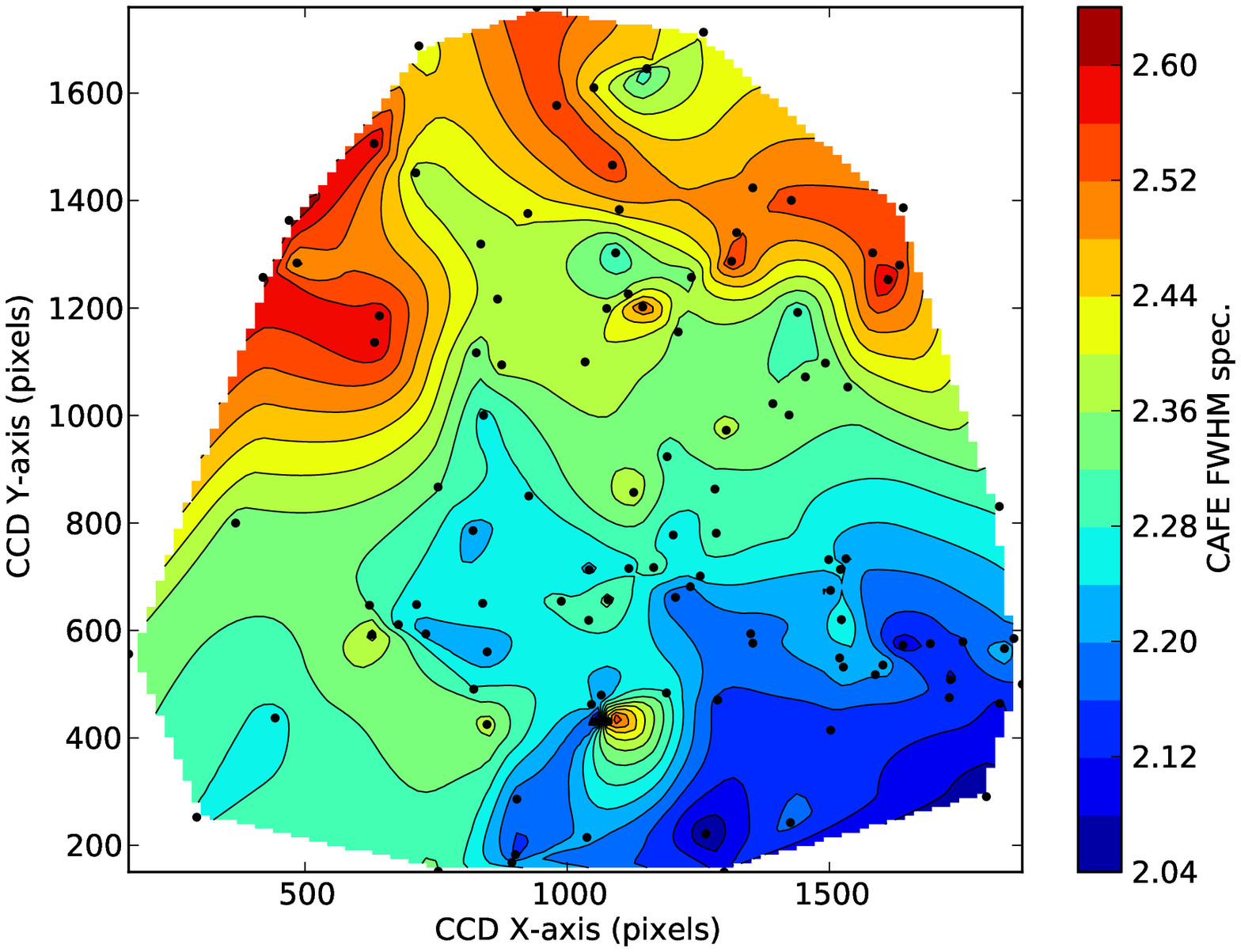}
\includegraphics[width=\hsize]{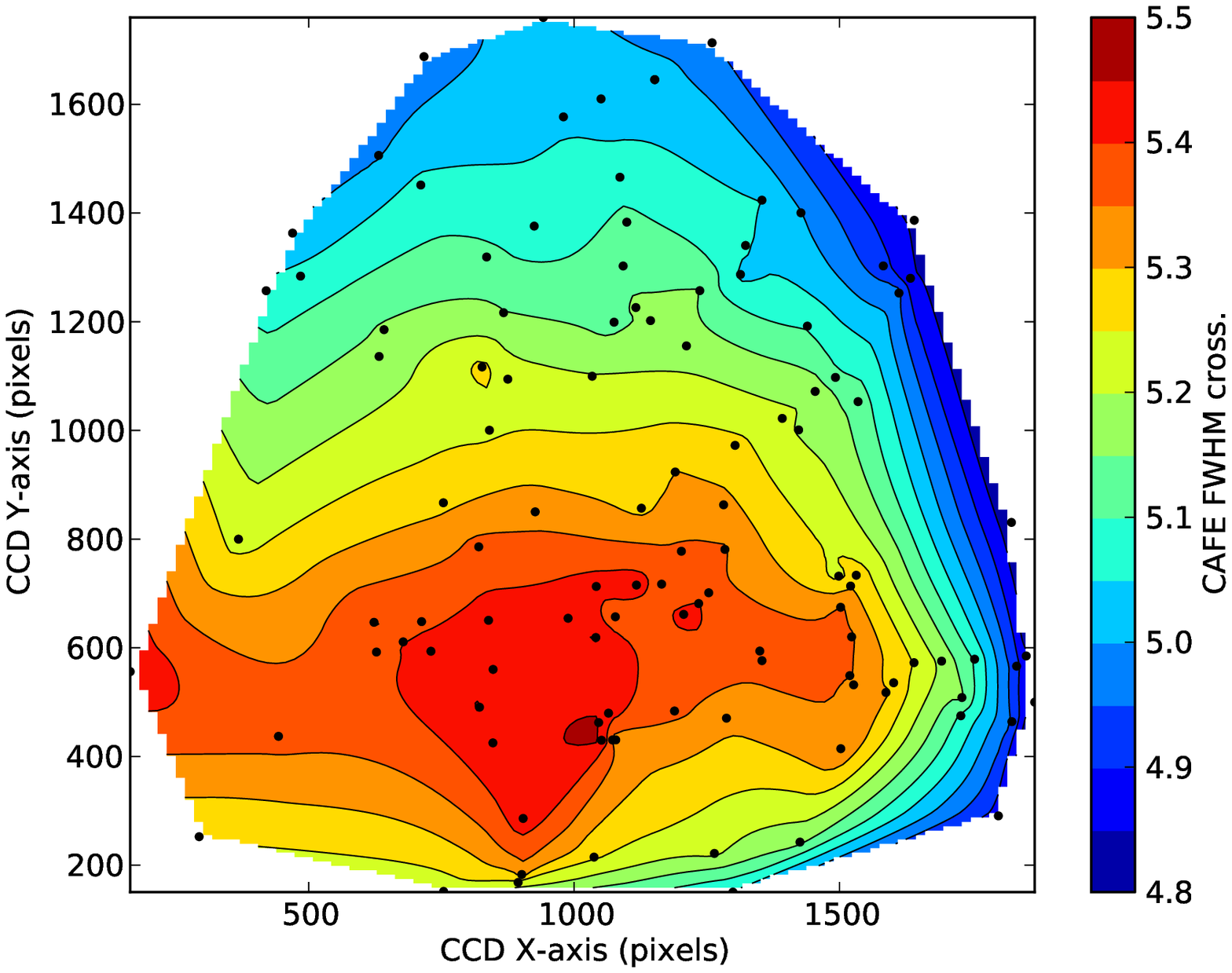}
\caption{CAFE: FWHM of the arc-lamp emission lines. Upper panel
shows the distribution of the FWHM of the arc-lamp emission lines
along the dispersion axis (X-axis), across the field-of-view of
the CCD. Lower panel shows a similar distribution for the FWHM
along the cross-dispersion axis.\label{fig:fwhm}}
\end{center}
\end{figure}

One of the main goals in the design and manufacturing of CAFE was
to achieve a better resolution than the one of its predecessor,
with a better efficiency and stability. CAFE was designed to
achieve a maximum resolution of $R\sim$70000 (\cite{san2007}), in the optimal situation.

The resolution of the instrument is defined as the ratio between
the wavelength $\lambda$ and the minimum range of wavelengths that
can be resolved $\Delta\lambda$. In practice, $\Delta\lambda$ is
derived from the FWHM, in the spectral direction, of the emission
lines of arc-lamps. Assuming that these lines are (in general),
unresolved, the FWHM measures the instrumental resolution, i.e.,
the minimum wavelength elements to be resolved. Early measurements
in the laboratory indicate that the FWHM of these emission lines
were around $\sim$2.2 pixels, which mostly corresponds to a
resolution R$\sim$67000 at the average wavelength sampled by the
instrument ($\lambda$6500\AA).

A detailed derivation of the spectral resolution can be done after
a proper identification of the orders, the wavelength range
covered by each of them, and the corresponding sampling ratio per
pixel. Once reduced an arc-lamp exposure, using the pipeline
described latter, each of the 336 identified arc emission lines
are fitted with a one dimensional Gaussian function in both the
dispersion and cross-dispersion directions, deriving the FWHM in
both axis. The FWHM in the cross-dispersion axis illustrates how
well each order is separated form the adjacent ones, and by which
amount they are contaminated by cross-talk. Fig.\ref{fig:fwhm}
shows the distribution of both FWHMs across the field-of-view of
the CCD.

On the other hand, the FWHM in the dispersion axis is a direct
measurement of the spectral resolution. Once it is derived the
FWHM of each of the identified lines, a clipping algorithm rejects
those values ($<10\%$) that deviated more than 3$\sigma$ of the
mean one. The derived FWHM is multiplied by the step in wavelength
per pixel and divided to the wavelength of the line, to derive the
instrumental resolution ($R$).

\begin{figure}
\begin{center}
\includegraphics[angle=270, width=\hsize]{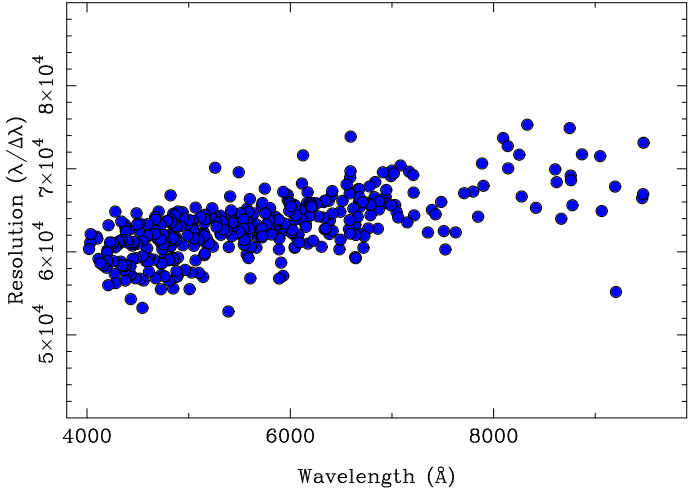}
\caption{CAFE: Wavelength resolution derived from the estimation
of the FWHM of the identified ARC emission lines in the dispersion
axis.\label{fig:res}}
\end{center}
\end{figure}

Fig.\ref{fig:res} shows the distribution of the spectral
resolution along the wavelength derived from the first ThAr lamp
observed the night of the 16th of June 2011. Similar distributions
are found for any of the arc calibration frames taken along the
commissioning run. In average, the instrumental resolution
estimated on real data corresponds to $\sim$63000 $\pm$ 4000.
There is a clear trend in the resolution from the blue to the red
range, with the resolution being $\sim$60000 in the blue end
($\sim$4000\AA), and about $\sim$70000 in the red end
($\sim$9500\AA). This median value is statistically dominated by
the values at wavelengths bluer than $\sim$5500\AA, where there
are more identified arc lines. As anticipated, the resolution at
the average wavelength of $\sim 6500$\AA\ is $\sim$ 65000.

Following the specifications of the design, CAFE was built and
calibrated to produce a similar accurate image quality at any
position across the CCD. I.e., it was a goal of the design to have
the same FWHM in the ThAr spots from blue to red arm at any order,
at least in the spectral axis. A constant FWHM (in pixels)
produces an unavoidable change in the resolution along the
spectral range, as the one observed here.

This resolution is, in average, larger than the one that could be
achieved by
FOCES\footnote{http//www.caha.es/pedraz/Foces/spec\_resol.html}.
With that instrument, it was feasible to reach a resolution of
about $\sim$65000 only when observing with the narrowest slit
width (with the consequent loss of signal-to-noise), and it was
not feasible to achieve a better resolution above this value.

\subsection{Stability of the Focus}

CAFE was designed to stabilize the camera focus (and therefore the
resolution) as much as possible. Due to that, compared with its
predecesor, many moving elements have been replaced by fixed ones.
To test if this goal has been achieved it is required to repeat
the measurements described in Section\ref{sec:focus} for
different ARC-lamps exposures taken under different conditions.

\begin{figure}
\begin{center}
\includegraphics[angle=270, width=\hsize]{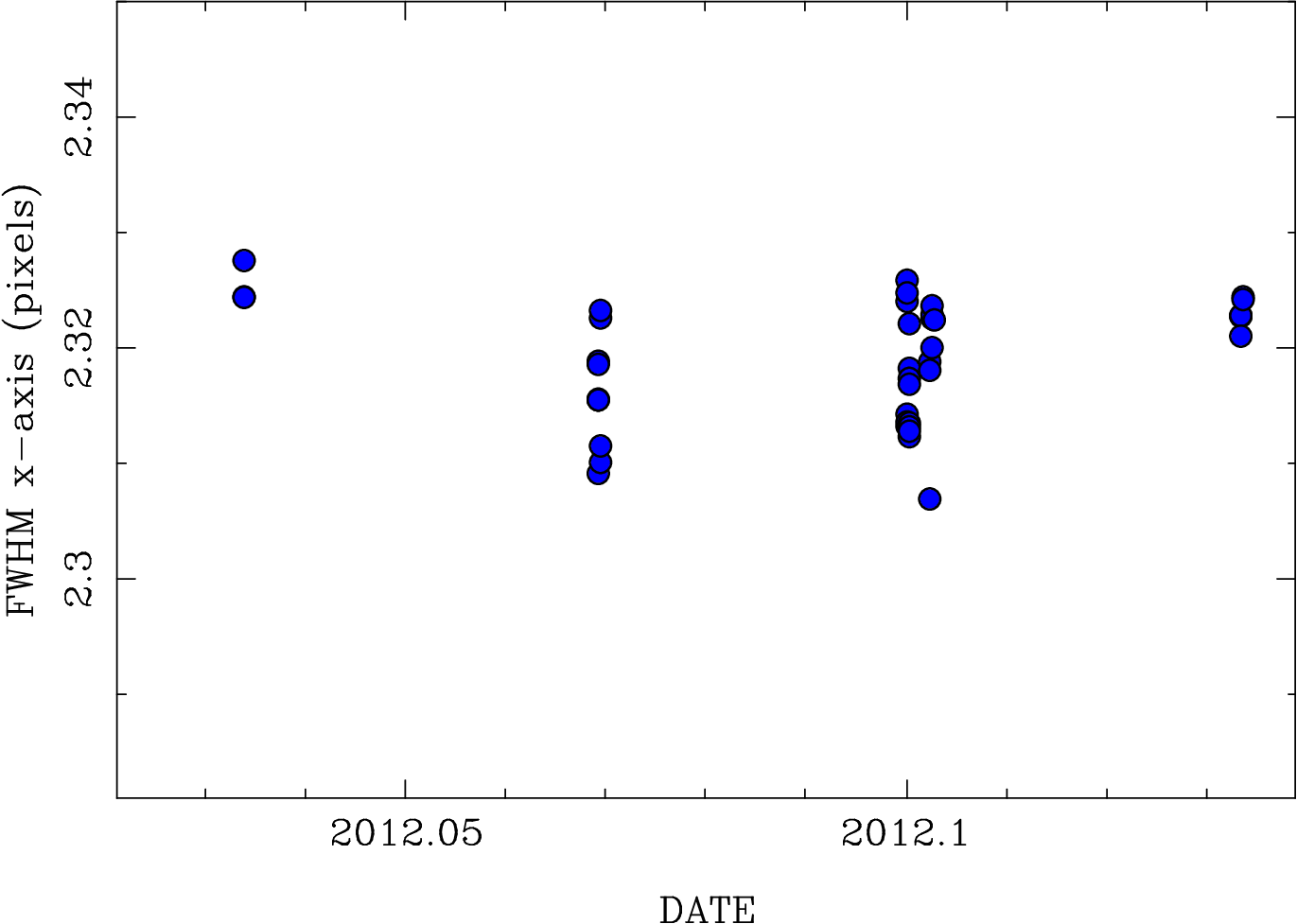}
\includegraphics[angle=270, width=\hsize]{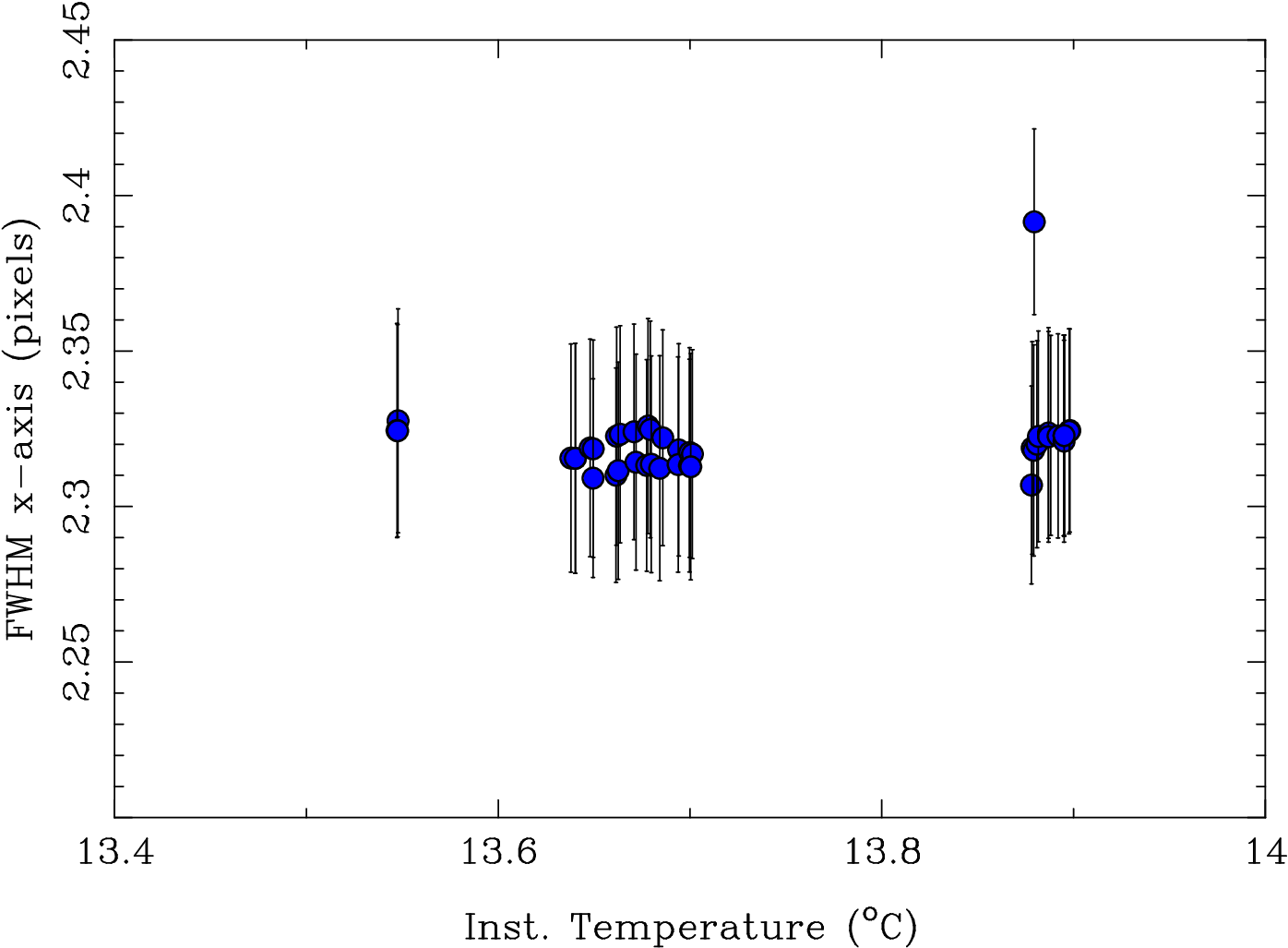}
\caption{CAFE: Stability of the spectral resolution. Upper panel
shows the distribution of the average FWHM of the ARC emission
lines along the dispersion axis (e.g., the one shown in Fig.\ref{fig:res}), for the different ARC frames taken along the
commissioning run. Lower panel shows a similar distribution along
the internal temperature of the instrument, measured with the four
sensors described in Section 3.1. \label{fig:sta}}
\end{center}
\end{figure}

So far, we got 34 ThAr ARC lamp frames along the commissioning
run. We repeated the procedure described before for each of them,
deriving the mean (and standard deviation) of the FWHMs values
measured for each of them, and compacted one each other. Fig.\ref{fig:sta} shows the distribution of these mean FWHMs along the
time (upper panel) and along the internal temperature of the
instrument (lower panel). It is important to note here that although the
instrument is equipped with a thermal controlling system, this
system was not operational during the Commissioning run.
Therefore, any effect of the temperature on the focus (and the
stability of the resolution) should be detected in this plot.

\begin{figure*}
\begin{center}
\includegraphics[height=6cm]{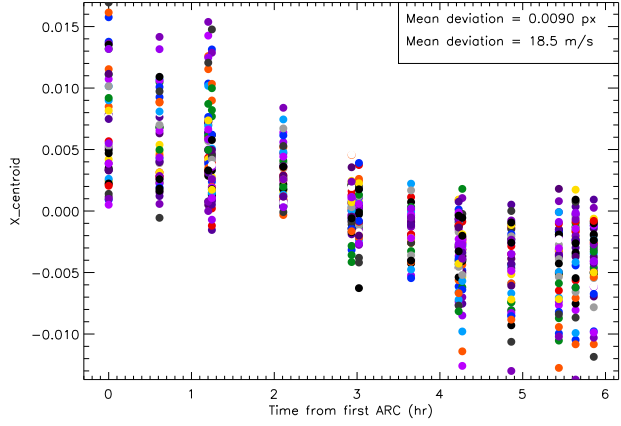}
\caption{Stability of CAFE along a 6 hours observing night on June
23rd, 2012. The mean dispersion of the centroid position in the
X-axis for the different Th-Ar frames are shown.43 spots, represented with
different colors, have been used in 14 ThAr arc images to generate this figure.
 \label{fig:stability}}
\end{center}
\end{figure*}

The average value of the FWHMs along the dispersion axis range
yields between 2.3-2.35 pixels, without any significant variation
either with the time and/or the temperature. The variation across
the field is much larger (see Fig. \ref{fig:fwhm}) than any of
possible detected variation along the time/temperature. In this
regards, the goals of the design have been completely full-filled.

It is required to (1) do this analysis for any of the observed
nights here after to feed the statistics with more data, and (2)
repeat the analysis if the focus is redone.

Fig.\ref{fig:stability} shows the high stability of CAFE along a
6 hours observing night on June 23rd, 2012. We used 43 spots in 14
ThAr arc images along the night to follow their centroid values
(in the X and Y directions). The mean dispersion of the centroid
position in the X-axis for the different images is 0.009 pixels
while 0.010 pixels is found for the Y direction, resulting in a
radial velocity precision of around 18.5 m/s and 21.2 m/s,
respectively. It is also important to note a slight dependence on
the centroid position along the night. However, it can be easily
corrected by using the closest ThAr arc to wavelength calibrate
the science images and hence achieve the mentioned precisions. We
detect different trends depending on each particular night so, if
the scientific program requires high precision radial velocity
measurements, we would strongly recommend to obtain arc
calibrations prior and after each science image.

\begin{figure*}
\begin{center}
\includegraphics[height=6cm]{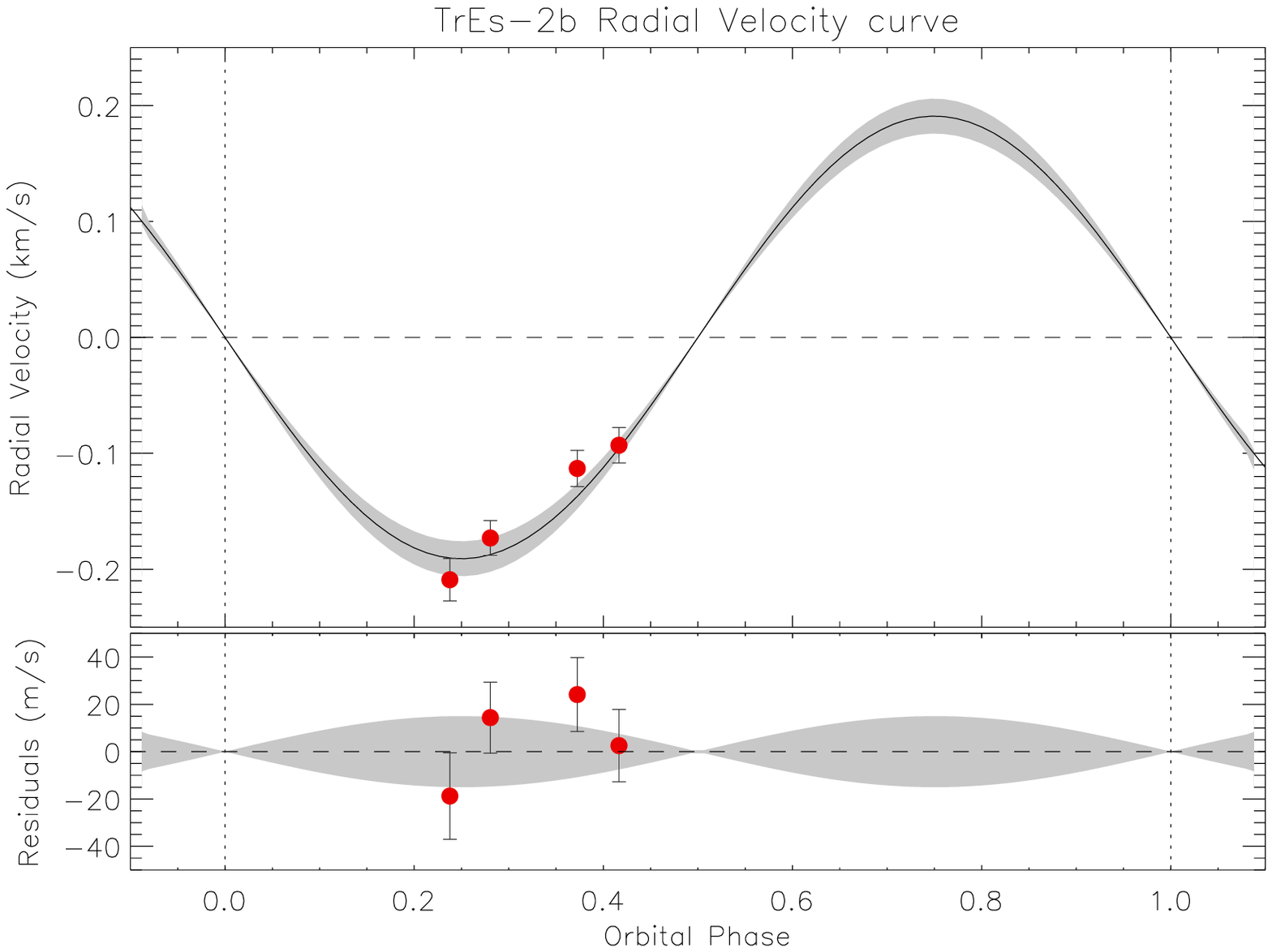}
\includegraphics[height=6cm]{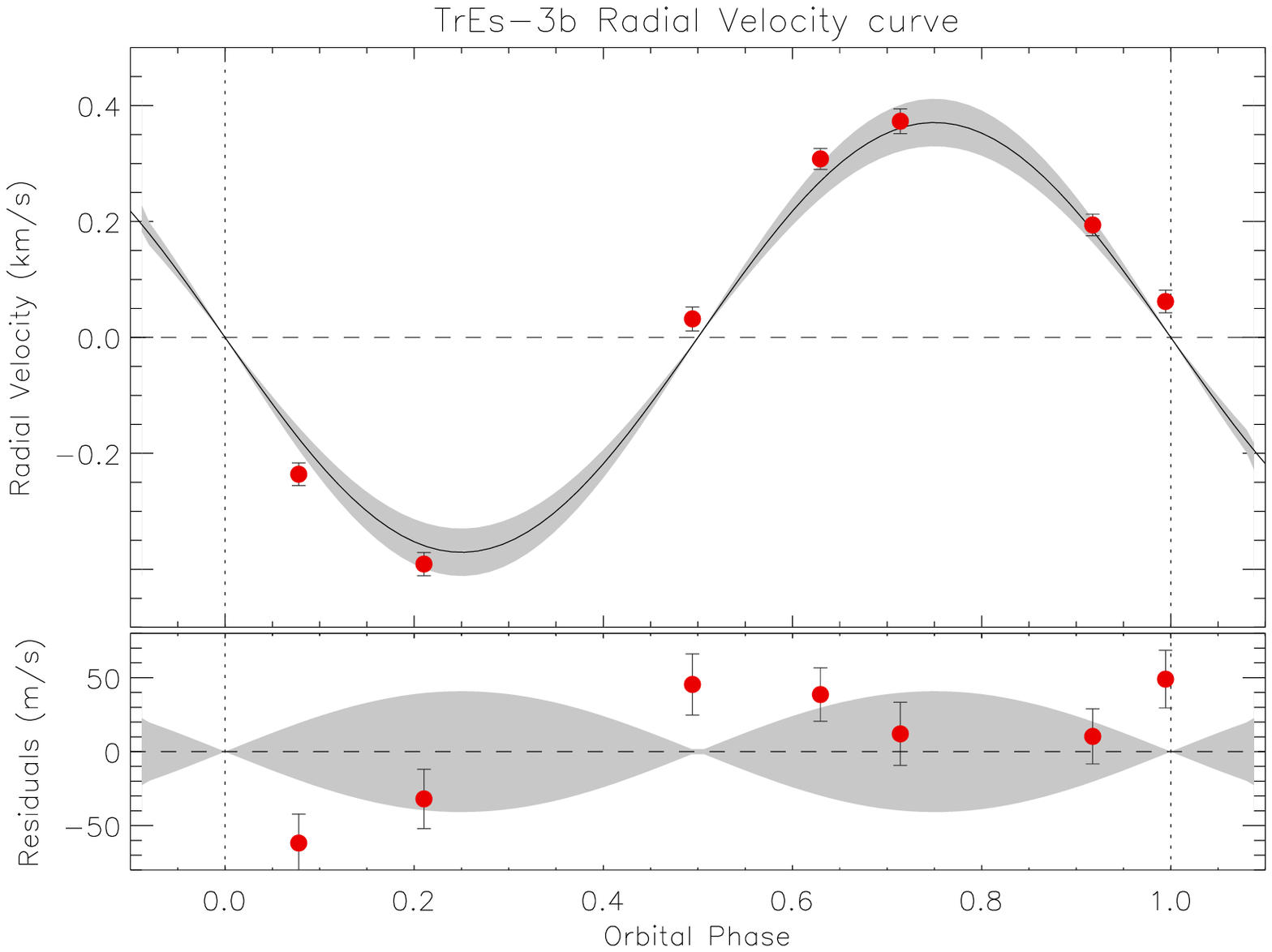}
\caption{Radial velocity curves for two well-known extra solar
planets, TrEs-2b ({\it left-panel}), and TrEs-3b ({\it
right-panel}), derived from the early measurements taken during
the CAFE commissioning run. Red circles represent the derived
values for the radial velocity. The black solid line represents
the theoretical curve assuming the simple expression:
$v_r=\frac{2\pi a}{P} sin(i) \frac{M_p}{M_p+M_s} sin(\phi)$ where
a is the semi-major axis, P is the orbital period, $i$ is the
orbital inclination and $\phi$ is the orbital phase. The shaded
region has been calculated by error propagation of the published
values in the previous expression. The lower panel shows the
residuals for the fit. \label{fig:TrEs}}
\end{center}
\end{figure*}

\begin{table}
\caption{Summary of the Observed objects along the Commissioning
nights}
\label{tab:SN}      
\begin{center}
\begin{tabular}{rlrrr}        
\hline\hline                 
Date & Object & Exposure & AB    & S/N  \\
     & Name   & Time     & (mag) &      \\
\hline
11-06-15 22:30:10 & HAT-P-12b  & 2700 & 12.4 & 57.3 \\
11-06-15 23:22:36 & WASP-24b   & 1800 & 11.3 & 67.7 \\
11-06-16 00:29:44 & KOI-561B   & 3600 & 13.6 & 45.3 \\
11-06-16 00:45:04 & HD154345   & 30   &  6.8 & 56.3 \\
11-06-16 00:49:41 & HD154345   & 60   &  6.3 & 48.3 \\
11-06-16 01:27:21 & TrES-2b     & 1800 & 11.5 & 75.4 \\
11-06-16 02:24:05 & Tres-3b    & 2700 & 12.2 & 70.1 \\
11-06-16 02:57:13 & TrES-2b     & 1800 & 11.3 & 75.9 \\
11-06-16 03:03:00 & THD182572  & 30   &  4.7 & 62.5 \\
11-06-17 21:13:45 & HZ44       & 1800 & 11.8 & 53.0 \\
11-06-17 21:46:51 & Feige66    & 1200 & 10.8 & 62.5 \\
11-06-17 22:14:10 & BD+33d2642 & 1200 & 11.4 & 56.2 \\
11-06-17 23:14:34 & TrES-3b    & 2700 & 12.6 & 46.7 \\
11-06-17 23:19:53 & HD115404   & 30   &  6.7 & 53.4 \\
11-06-17 23:35:51 & HD139323   & 30   & 7.20 & 35.1 \\
11-06-18 00:11:11 & P330D      & 1800 & 12.9 & 37.7 \\
11-06-18 00:52:22 & TrES-2b    & 1800 & 11.4 & 42.7 \\
11-06-18 01:04:29 & HD151541   & 50   &  7.2 & 45.8 \\
11-06-18 01:33:24 & BD+25d4655 & 600  & 10.6 & 34.1 \\
11-06-18 02:32:04 & TrES-3b    & 2700 & 12.2 & 57.5 \\
11-06-18 03:12:31 & TrES-2b    & 1800 & 11.0 & 42.5 \\
11-06-18 03:23:20 & HD090404   & 50   &  7.1 & 64.6 \\
11-06-18 20:58:34 & P330D      & 1800 & 13.2 & 33.4 \\
11-06-18 21:04:44 & HD115404   & 30   &  7.1 & 31.3 \\
11-06-18 22:03:56 & TrES-3b    & 2700 & 12.3 & 58.2 \\
11-06-18 23:00:47 & Kepler-561 & 2700 & 13.8 & 30.0 \\
11-06-18 23:35:28 & TrES-2b    & 1800 & 11.3 & 61.4 \\
11-06-18 23:44:03 & HD139323   & 30   &  7.3 & 46.7 \\
11-06-19 00:41:05 & TrES-3b    & 2700 & 12.0 & 58.5 \\
11-06-19 01:30:06 & Kepler-561 & 2700 & 13.6 & 32.3 \\
11-06-19 02:08:19 & TrES-2b    & 1800 & 11.0 & 61.2 \\
11-06-19 02:18:33 & HD151541   & 50   &  7.1 & 52.3 \\
11-06-19 03:12:25 & Kepler-561 & 2700 & 14.5 & 20.0 \\
11-06-19 03:25:07 & HD190404   & 50   &  6.8 & 62.1 \\
\hline
\end{tabular}
\end{center}
\end{table}

\subsection{Signal-to-Noise}

CAFE was designed to be more efficient than FOCES. For doing so
new branch fibers, optics, higher-efficiency elements and less
movable elements were included.

To determine if we have achieved this goal, we used the
flux-calibrated spectra of the different objects observed during
the commissioning run to derive the average S/N ratio (per
spectral pixel). For doing so, an automatic procedure was included
in the CAFE pipeline, that compares the extracted signal, with the
derived variance after propagating the different reduction steps.

\begin{figure}
\begin{center}
\includegraphics[width=\hsize]{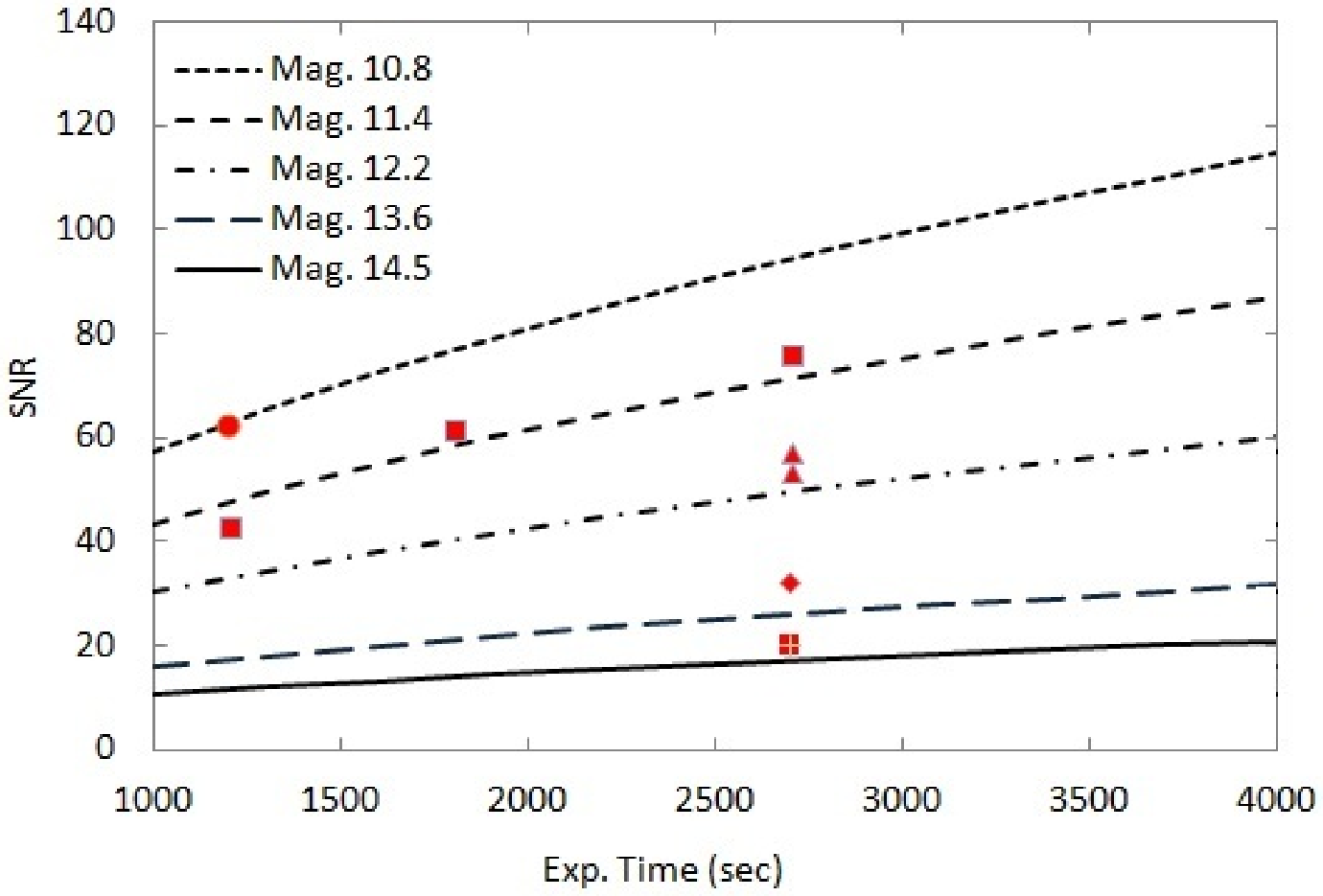}
\includegraphics[width=\hsize]{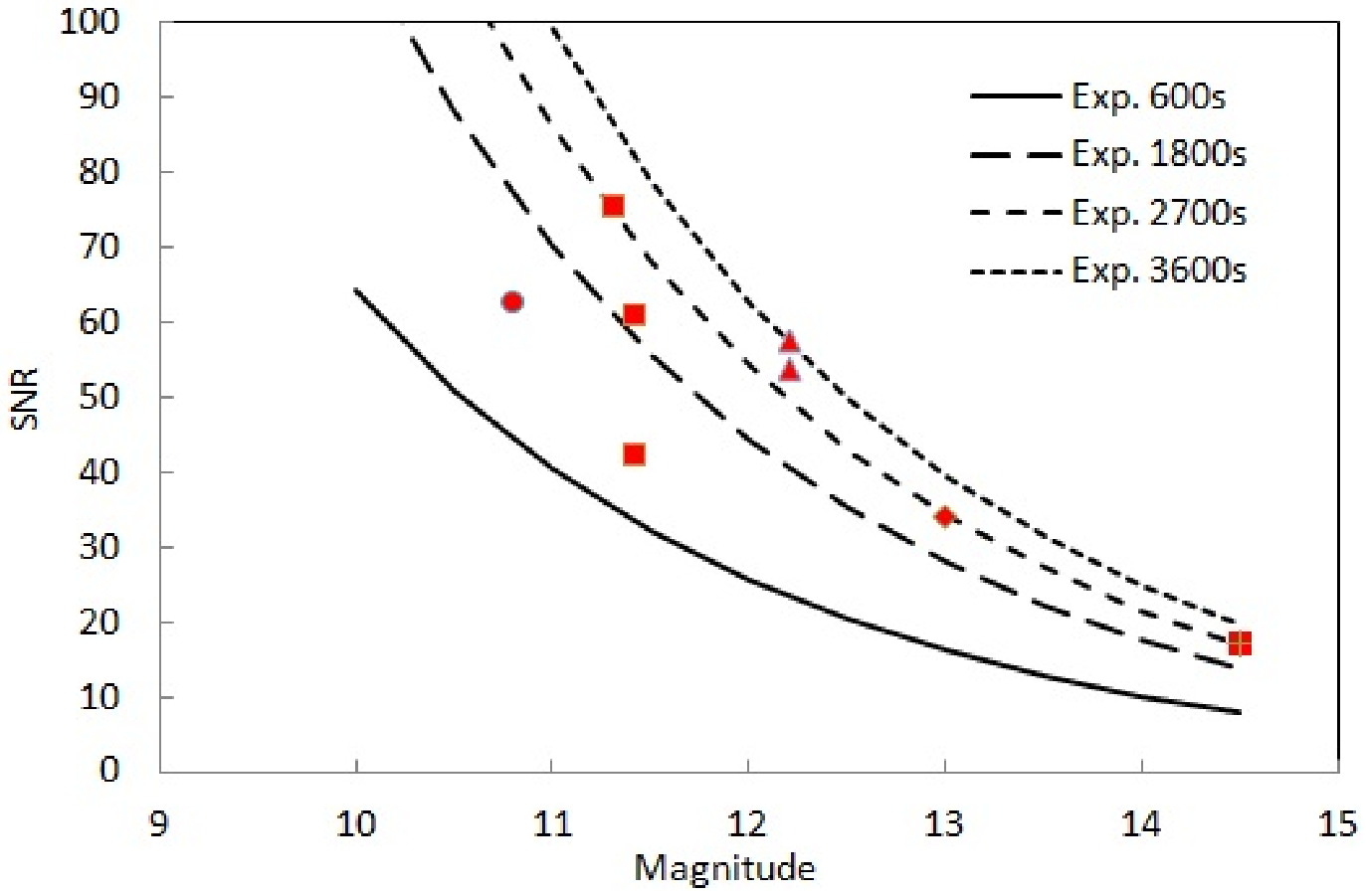}
\caption{Upper panel: The plot shows the SNR as a function of time for a
  given magnitude. Some experimental data for the corresponding magnitude have
  been overlapped to the theoretical curves for which magnitudes were derived. Lower
  panel: SNR is shown as a fucntion of magnitude for different fixed exposure
  time. Same experimental data than in the upper case have been plotted to
  be consistent. \label{fig:SN}}
\end{center}
\end{figure}

Fig.\ref{fig:SN} shows the theoretical curves derived from a exposure time
calculator for CAFE. The achieved SNR as a function of time for a given
magnitude has been included for the corresponding theoretical curve.

Table \ref{tab:SN} shows the results of this S/N analysis, for all
the targets observed along the commissioning run, including the
date, the name of the target, the $V$-band magnitude and the S/N
at the average wavelength of this band ($\sim$5500\AA). However, a similar
S/N, within a 20\% is derived for the full wavelenght range between
5000-6500\AA, depending more on the shape of the continuum of the considered
target (e.g stellar type) than in the properties of the instrument. These
results can be directly compared with the ones derived for
FOCES\footnote{http://www.caha.es/pedraz/Foces/signal.html}. We
highlight here the results derived for the G5 star with
$V$-band magnitude of 6.9 mag, derived with FOCES, obtaining a
S/N$\sim$41 for a exposure time of 60 sec. This can be compared
with the result we obtain for HF151541, with CAFE, a $V\sim$7.1
mag star, for which we obtain a S/N$\sim$68 with a exposure time
of 60 sec. The faintest object listed in the FOCES reference
web-page, is a 0p star with a luminosity of $V\sim$10.5 mag,
for which it was obtained a spectra with a S/N$\sim$25 with a
exposure time of 600 seconds. A similar star, BD+25d4655,
V$\sim$10.6 mag, was observed with CAFE, for which we obtained a
spectrum with a S/N$\sim$45, for with a similar exposure time
($t_{exp}\sim$ 600 sec).

In average, the S/N derived by CAFE is twice larger than the one
derived with FOCES for targets with similar luminosity and using
similar exposure times.

\begin{figure*}
\begin{center}
\includegraphics[width=7cm]{aceitunoFig17.ps}
\includegraphics[width=10.5cm, clip=true,trim=1 1 10 40]{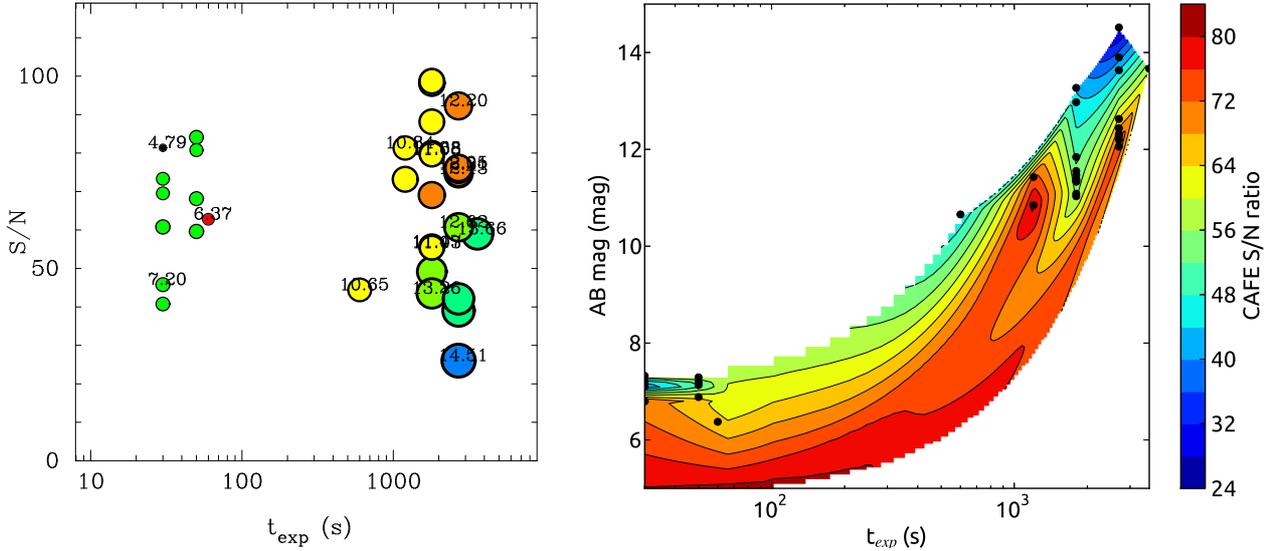}
\caption{CAFE: Results from the analysis of the S/N. Left panel
shows the distribution of the S/N along the exposure time, for the
different observed objects. Color/size of the plotted symbols
indicate the brightness of the considered object. Right panel
shows the S/N distribution as a function of the brightness and the
exposure time. \label{fig:SNmap}}
\end{center}
\end{figure*}

Fig.\ref{fig:SNmap} summarizes the results of this analysis,
showing in two different representations the dependency of the S/N
with the intrinsic brightness of the observed object and the
adopted exposure time. Based on these results, the limiting
magnitude of CAFE would be $\sim$15 mag, for an exposure time of 1
hour, with a S/N ratio of $\sim$20. We consider that the goal of
providing an instrument more efficient than FOCES has been
fulfilled.

These figures/tables should be fed with any additional information
provided in any further observing run to derive much more accurate
results/expectations.

\subsection{Radial velocities measurements}

Fig.\ref{fig:stability} shows the high stability of CAFE along a
6 hours observing night on June 23rd, 2012. We used 43 spots in 14
ThAr arc images along the night to follow their centroid values
(in the X and Y directions). The mean dispersion of the centroid
position in the X-axis for the different images is 0.009 pixels
while 0.010 pixels is found for the Y direction, resulting in a
radial velocity precision of around 18.5 m/s and 21.2 m/s,
respectively. It is also important to note a slight dependence on
the centroid position along the night. However, it can be easily
corrected by using the closest ThAr arc to wavelength calibrate
the science images and hence achieve the mentioned precisions. We
detect different trends depending on each particular night so, if
the scientific program requires high precision radial velocity
measurements, we would strongly recommend to obtain arc
calibrations prior and after each science image.

\begin{figure*}
\begin{center}
\includegraphics[height=6cm]{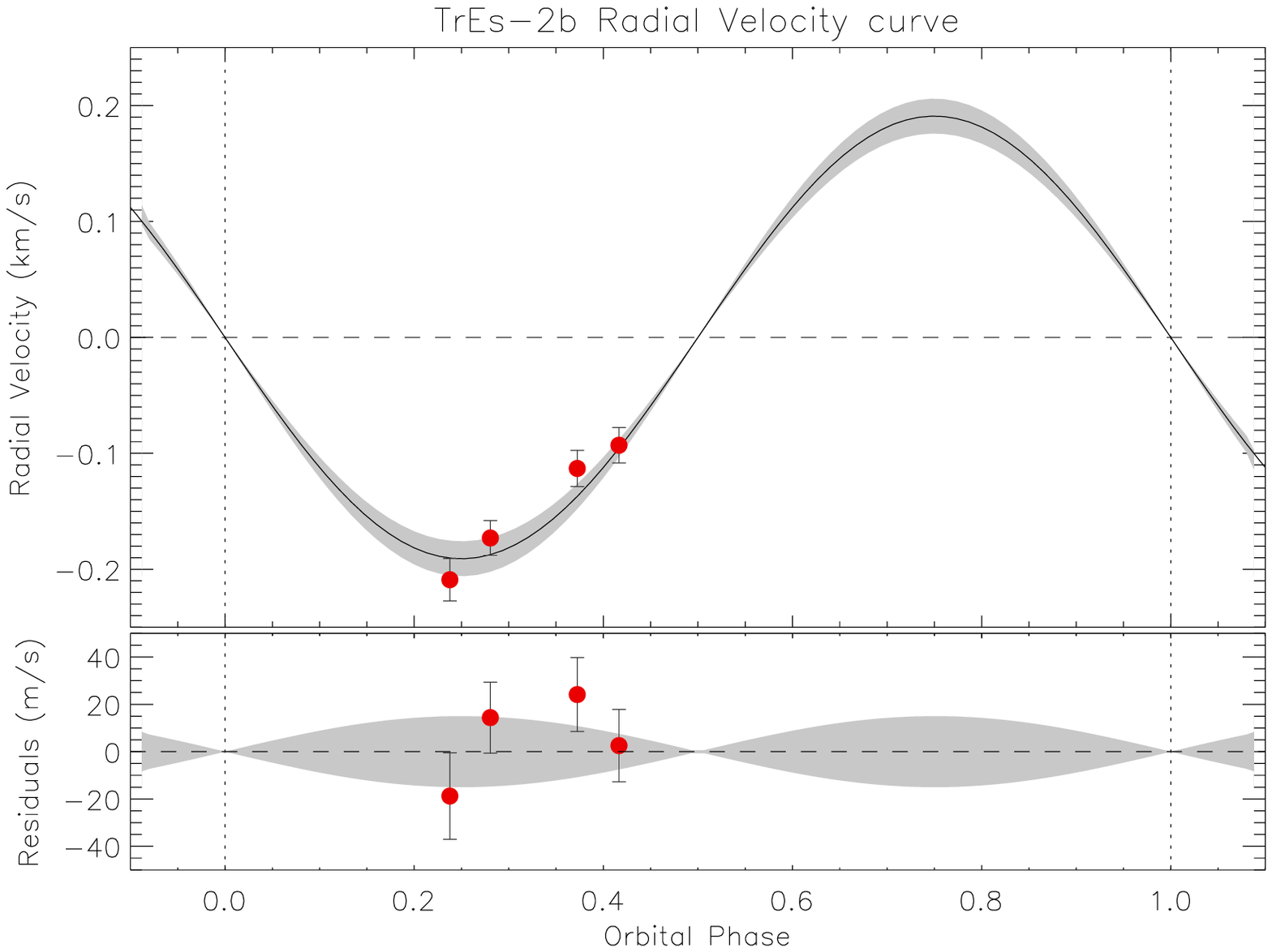}
\includegraphics[height=6cm]{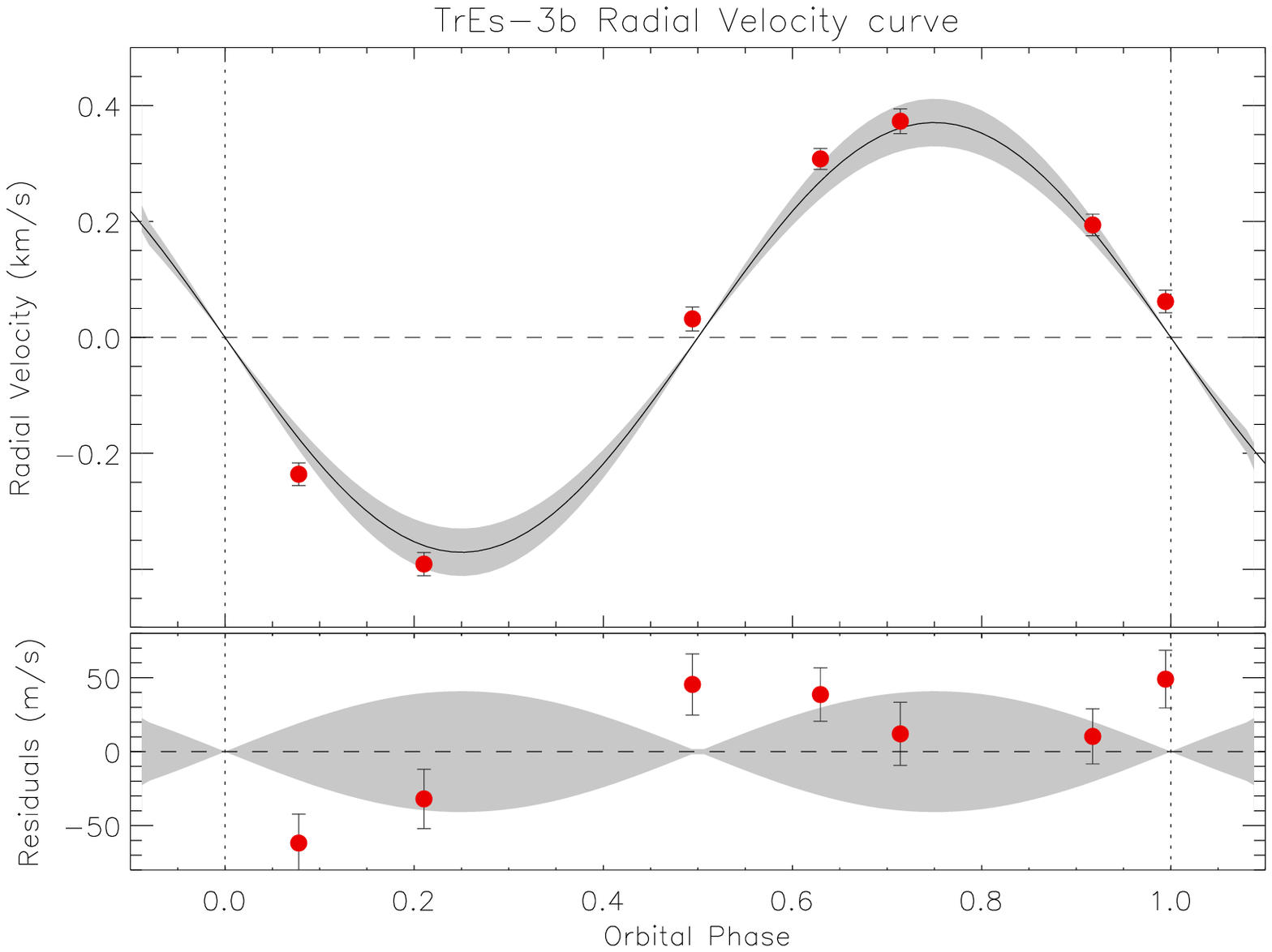}
\caption{Radial velocity curves for two well-known extra solar
planets, TrEs-2b ({\it left-panel}), and TrEs-3b ({\it
right-panel}), derived from the early measurements taken during
the CAFE commissioning run. Red circles represent the derived
values for the radial velocity. The black solid line represents
the theoretical curve assuming the simple expression:
$v_r=\frac{2\pi a}{P} sin(i) \frac{M_p}{M_p+M_s} sin(\phi)$ where
a is the semi-major axis, P is the orbital period, $i$ is the
orbital inclination and $\phi$ is the orbital phase. The shaded
region has been calculated by error propagation of the published
values in the previous expression. The lower panel shows the
residuals for the fit. \label{fig:TrEs}}
\end{center}
\end{figure*}

As it was said in previous sections, during the Commissioning run,
we observed some already known planets in order to test the CAFE
capabilities with real data. Spectra of parent stars TrEs-3
\cite{don2007} and TrEs-2 \cite{don2006} were acquired in
different phases of their orbits.

The former system is composed by a $M_psin(i)=1.91^{+0.06}_{-0.08}
M_J$ planet orbiting a G-type star (TrES-3) of
$M_s=0.924_{−0.04}^{+0.012} M_{\odot}$. The physical and
dynamical characteristics of this system causes a movement of the
parent star around the center of masses with an amplitude of
$K=371\pm 41$ m/s (error calculated during the orbit's
quadrature). In Fig\ref{fig:TrEs}, we show the observational
points taken with CAFE after reducing the raw data with the
pipeline explained in the previous section. Radial velocity for
each spectra has been derived by cross-correlating the 50 orders
with the highest signal to noise ratio with a solar
spectrum\footnote{www.bass2000.obspm.fr}. When combining all the
cross-correlation functions of the orders, we weigthtened them
according to their FWHM. In the mentioned figure, we also plot the
radial velocity curve according to the published parameters of the
system\footnote{Obtained from the www.exoplanet.eu website}.A nice
fit with the largest residuals being of the order of 50 m/s is
obtained.

We proceed in the same manner for the second star, TrES-2, that
hosted a planet lighter than TrES-3b. This planet is less massive
than TrES-3b ($M_psin(i)=1.253\pm 0.052 M_J$) and orbits a heavier
star of $M_s=0.980\pm 0.062M\odot$. Hence the amplitude of the
radial velocity curve is significantly smaller, $K=190\pm 15$ m/s.
Although, in this case, a slightly poor sampling of this curve was
taken, we can see in Fig\ref{fig:stability} that smaller residuals
were obtained. Again, observed points with CAFE fit what is
expected for this system, with residuals smaller than 20 m/s.

\begin{figure*}
\begin{center}
\includegraphics[height=6cm]{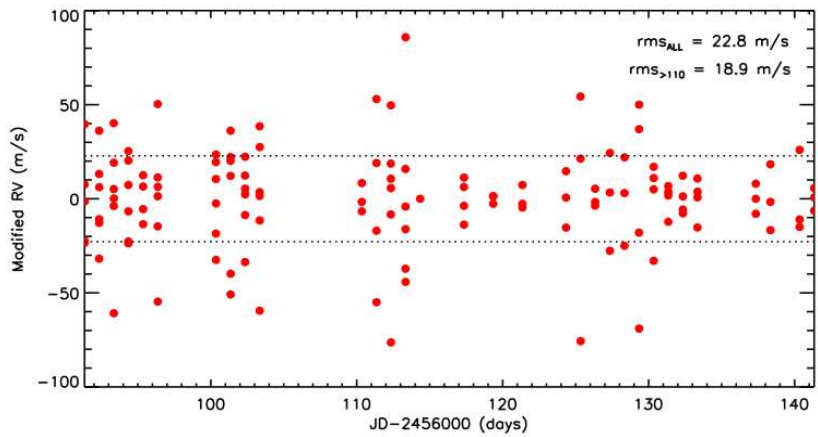}
\caption{ RMS error of radial velocity measurements of the standard HD124292.  
 \label{fig:rmserror}}
\end{center}
\end{figure*}

According to our observations of radial velocity standards, we find precisions
of 22 m/s in the case of HD124292 if we remove the night-by-night trend
(see Fig\ref{fig:rmserror}) .This trend could be cause by different effects such as stellar pulsation or
night-by-night instrumental effects due to some small problems releated to the
arc lamp intensity stability. This is due basically to the optimal temperature required for 
 the lamp to produce sharp emission lines, and not broader ones, which
 centroids are different. Only when the lamp is switched-on during the 
 complete nights it is reached an optimal stability in the position
 of the arc-lamps. However, this is not feasible, due to the possible
 effects in straylight. 

Note that these radial velocity
measurements  have been determined by using high signal-to-noise spectra
(greater than 80) of bright known standards and cross-correlating it with
synthetic spectra of the same spectral type. The larger rms obtained in the
case of TrEs-3b could be due among others to the fact that the Tres-3 spectra have lower signal-to-noise ratio (around 30).

\section{Summary and Conclusions}
\label{sum}

In this paper, we have presented the design, manufacturing and
performance analysis of CAFE. We showed that the instrument was
built according to the demanding requirements which resulted in an
instrument of excellent stability and efficiency. First tests at
the telescope were presented and lead to very encouraging results,
which might be summarized as follows:

\begin{itemize}

\item The instrument is fully operational and publically
accessible.

\item The resolution estimated on real data corresponds to
62000$\pm$5000A.

\item Based on real observations, the limiting magnitude of CAFE
would be $\sim$15 mag, for an exposure time of 1 hour, with a S/N
ratio of $\sim$20.

\item Two known planets have been observed and their radial
velocities measured to test the CAFE capabilities.

\item The high stability of the instrument shows that a radial
velocity precision about $\sim$20$m~s^{-1}$ might be achieved.

\end{itemize}

\section{Acknowledgements}

  The authors thank the sub-programs of {\it Viabilidad, Dise\~no, Acceso
  y Mejora de ICTS} , ICTS-2008-24 and ICTS-2009-32, the {\it PAI
  Proyecto de Excelencia} P08-FWM-04319 and the funds of the PAI research group
  FQM360, and the MICINN program AYA2010-21161-C02-02, AYA2010-22111-C03-03, CDS2006-00070 and PRICIT-S2009/ESP-1496. We are
  grateful to all the Calar Alto staff and in particular to E. de Guindos, E. de Juan for the computer support, to S. Reinhart, M.
  Pineda , A. Garcia and J. Garcia who worked out during the mechanical
  assembling; to N. Vico, L. Hernandez, J. F. Lopez and J. Marin for their
  contributions to the new electronic controller and V. Gomez, M.
  Aguila, R. Lopez who their contributions to the preparation of
  the room that holds the instrument.We also want to thanks to the referee,
  which comments have improved this manuscript.

\bibliographystyle{aa}
\bibliography{CAFEI}


\end{document}